\documentclass[aip,jcp,reprint,showkeys]{revtex4-1}
\usepackage{bm,graphicx,tabularx,array,dcolumn,xcolor,microtype,multirow,amscd,amsmath,amssymb,amsfonts,physics,siunitx}
\usepackage[version=4]{mhchem}
\usepackage[utf8]{inputenc}
\usepackage[T1]{fontenc}
\usepackage{txfonts}
\usepackage{simpler-wick}

\usepackage[normalem]{ulem}
\definecolor{hughgreen}{RGB}{0, 128, 0}

\newcommand{\ie}{i.e.}

\newcommand{\SupInt}{\textcolor{blue}{Supplementary Material}}

\usepackage[
	colorlinks=true,
    citecolor=blue,
    linkcolor=blue,
    filecolor=blue,      
    urlcolor=blue,	
    breaklinks=true
	]{hyperref}
\newcommand{\br}{\bm{r}}
\newcommand{\bx}{\bm{x}}

\newcommand{\RHF}{\text{RHF}}
\newcommand{\UHF}{\text{UHF}}
\newcommand{\GHF}{\text{GHF}}
\newcommand{\s}{\text{s}}
\newcommand{\tL}{L}
\newcommand{\tR}{R}

\newcommand{\Rc}{R_{\text{c}}}

\newcommand{\cS}{\mathcal{S}}
\newcommand{\cSx}{\cS_x}
\newcommand{\cSy}{\cS_y}
\newcommand{\cSz}{\cS_z}
\newcommand{\ms}{m_s}

\newcommand{\ta}{\theta^\alpha}
\newcommand{\tb}{\theta^\beta}

\renewcommand{\rho}{n}
\newcommand{\rhos}{\rho^{\sigma}}
\newcommand{\rhoa}{\rho^{\alpha}}
\newcommand{\rhob}{\rho^{\beta}}
\newcommand{\rhow}{\rho^{w}}
\newcommand{\rhoaa}{\rho^{\alpha \alpha}}
\newcommand{\rhoab}{\rho^{\alpha \beta}}
\newcommand{\rhoba}{\rho^{\beta \alpha}}
\newcommand{\rhobb}{\rho^{\beta \beta}}
\newcommand{\rhossp}{\rho^{\sigma \sigma'}}

\newcommand{\psis}{\psi^\sigma}
\newcommand{\psisp}{\psi^{\sigma'}}
\newcommand{\psia}{\psi^\alpha}
\newcommand{\psib}{\psi^\beta}

\newcommand{\tpsi}{\tilde{\psi}}
\newcommand{\psil}{\psi_\lambda}

\newcommand{\vext}{v_{\text{ext}}}
\newcommand{\ke}{T}

\newcommand{\UOX}{Physical and Theoretical Chemical Laboratory, Department of Chemistry, University of Oxford, Oxford, OX1 3QZ, U.K.}
\newcommand{\LCPQ}{Laboratoire de Chimie et Physique Quantiques, Universit\'{e} de Toulouse, CNRS, UPS, France}
\newcommand{\AIMMS}{Department of Chemistry and Pharmaceutical Sciences, Amsterdam Institute of Molecular and Life Sciences (AIMMS), Faculty of Science, Vrije Universiteit, De Boelelaan 1083, 1081HV Amsterdam, The Netherlands}

\begin{document}

%\title{\paola{The Hartree--Fock landscape of one-electron systems}}
%\title{{Landscape of fractional-spin Hartree--Fock for one electron}}
\title{Variations of the Hartree--Fock fractional-spin error for one electron}

\author{Hugh~G.~A.~\surname{Burton}}
\email{red.burton@chem.ox.ac.uk}
\affiliation{\UOX}
\author{Clotilde \surname{Marut}}
\affiliation{\LCPQ}
\author{Timothy J. \surname{Daas}}
\affiliation{\AIMMS}
\author{Paola \surname{Gori-Giorgi}}
\affiliation{\AIMMS}
\author{Pierre-Fran\c{c}ois \surname{Loos}}
\email{loos@irsamc.ups-tlse.fr}
\affiliation{\LCPQ}

\begin{abstract}
Fractional-spin errors are inherent in all current approximate density functionals, including Hartree--Fock theory, 
and their origin has been related to strong static correlation effects.
The conventional way to encode fractional-spin calculations is to construct an ensemble density that scales
between the high-spin and low-spin densities.
In this article, we explore the variation of the Hartree--Fock fractional-spin (or ghost-interaction) error in one-electron systems using restricted
and unrestricted ensemble densities, and the exact generalized Hartree--Fock representation.
By considering the hydrogen atom and \ce{H2^+} cation, 
we analyze how the unrestricted and generalized Hartree--Fock schemes minimize this error by localizing the electrons or rotating the spin coordinates.
We also reveal a clear similarity between the Coulomb hole of \ce{He}-like ions and the density depletion near the nucleus induced by the fractional-spin error in the unpolarized hydrogen atom.
Finally, we analyze the effect of the fractional-spin error on the M{\o}ller-Plesset adiabatic connection, excited states, and functional- and density-driven errors.
\end{abstract}

\maketitle
\raggedbottom
%%%%%%%%%%%%%%%%%%%%%%
\section{Introduction}
%%%%%%%%%%%%%%%%%%%%%%
 
When is Hartree--Fock (HF) theory exact for a one-electron system?
It is certainly exact for the ground-state of a single occupied
orbital containing a spin-up or spin-down electron (\ie, $\ms = \pm\frac{1}{2}$).
It must also be exact for any single non-collinear spin orbital, corresponding to a one-electron generalized HF wave function.\cite{Fukutome1981,Sykja1982,Calais1985,Lowdin1992,Mayer1993,HammesSchiffer1993,Stuber2003,Jimenez-Hoyos2011,Small2015a,Small2015b,Goings2015,Goings2018,Henderson2018,Jake2018}
But what about a one-electron problem with fractionally-occupied spin orbitals?
In this case, HF theory fails.
In fact, the resulting ``fractional-spin error'' is asymptotically equivalent to the static correlation error in stretched
diatomic molecules\cite{Cohen2008,Cohen2008b,MoriSanchez2009,Daas2020} and occurs in all standard density functional approximations.\cite{Cohen2008b,Mussard2017} 
Removing these errors requires more sophisticated functionals such as those based on the exact-exchange random-phase approximation\cite{Hesselmann2011,Erhard2016} or on the strictly-correlated-electrons (SCE) limit of density-functional theory (DFT).\cite{Chen2014,Vuckovic2015,Vuckovic2017}
This connection with static correlation has made the fractional-spin error a popular metric
for assessing and improving the quality of approximate density functionals, \cite{Cohen2008,MoriSanchez2009,Johnson2011,Cohen2012,Mussard2017,Su2018} 
many-body perturbation theory,\cite{Phillips2015}
and wave function methods.\cite{Steinmann2013,Cohen2009} 
Variants of fractional-spin DFT have also been successfully applied to various chemical problems, particularly for predicting the energetics of
open-shell diradical systems. \cite{Schipper1998,Filatov1999,Filatov2000a,Filatov2000b,Ess2011,Chai2012,Filatov2015}

The physical origins of the fractional-spin error have been extensively studied by Cohen, Mori-Sanchez and Yang.\cite{Cohen2008,Cohen2008b,MoriSanchez2009,Cohen2012}
In particular, they have shown that the energy of the exact functional with fractional spins should 
be constant.\cite{Cohen2008}
When an approximate functional fails to conserve this constancy {condition}, the resulting 
error can be associated with static correlation.
This condition is analogous to the piecewise linearity of the energy for fractional charges, which leads to delocalization errors when not satisfied.\cite{Cohen2008b,MoriSanchez2009}
Through fractional spins and fractional charges, many of the failures of electronic structure theory
can be studied as properties of the functional rather than the structure of an approximate 
density or wave function.\cite{Cohen2008b}
This analysis relies on the idea that a well separated fragment of a pure-state system might be locally described as an ensemble, 
and that it should be correctly described in this way by an energy functional (for an in-depth critical analysis, see Refs.~\onlinecite{Baerends2017,Baerends2020}).
Furthermore, these errors can be directly connected to the ghost interaction between different states in 
an ensemble density calculation.\cite{Gidopoulos2002}

While the relationship between fractional-spin errors and static correlation are now well established, 
the corresponding short-comings of the exchange-correlation functional remain elusive.
Hartree--Fock theory does not satisfy the fractional-spin condition, suggesting that {artificial Coulomb interactions} play a significant role.\cite{Cohen2008}
However, HF theory exists in many different forms depending on the symmetry restrictions applied
to the electronic state, and each of these approximations may yield different fractional-spin errors.
Restricted HF (RHF), uses the same spatial orbitals for both spin-up and spin-down electrons, 
conserving both $\cS^2$ and $\cSz$ spin symmetry.
The unrestricted HF (UHF) approach uses different orbitals for different spins, allowing broken $\cS^2$ symmetry but enforcing $\cSz$ symmetry.
{While UHF theory often correctly describes the dissociation of closed-shell molecules into open-shell fragments,
it yields qualitatively incorrect binding curves in systems such as \ce{H4^{2+}} (square and chain),\cite{MoriSanchez2014} 
\ce{F2},\cite{Gordon1987} or \ce{O2^{2+}},\cite{Nobes1991} and fails for processes where the total
$m_\text{s}$ value is not conserved as the nuclear coordinates change.\cite{Jimenez-Hoyos2011}}
Finally, generalized HF (GHF) allows each orbital to have high- and low-spin components and is 
not guaranteed to conserve either  $\cS^2$ or $\cSz$ symmetry.
Fukutome classified these three formalisms as time-reversal invariant closed-shell, axial spin density waves, and torsional spin density waves respectively,\cite{Fukutome1981} while extensions with complex orbitals also exist\cite{Pople1971,Ostlund1972,Small2015a,Goings2015,Hiscock2014,Burton2016,Burton2018} and
point group symmetry can always be independently imposed.

In this contribution, we systematically investigate the fractional spin error 
of the HF potential using different symmetry-restricted formalisms.
By considering the optimized HF energy and densities for a series of hydrogen-based one-electron models, we show that the fractional-spin error 
arises from an {artificial Coulomb interaction} %\cite{Perdew1981,Zhang1998} 
between the high-spin and 
low-spin densities that leads to artificially spin-polarized densities and additional mean-field solutions.
The magnitude of this error therefore depends on the form of HF approximation applied.
We also discover a density depletion zone created by this fractional-spin error in the \ce{H} atom that is
remarkably similar to the Coulomb hole (\ie, the difference between the exact and HF system-averaged pair densities) induced by
electron correlation effects in helium-like ions. \cite{Coulson1961,Pearson2009}
Finally, we demonstrate how the fractional-spin error decreases as the electron density becomes increasingly delocalized. 

Unless otherwise stated, all results are computed using Mathematica 12.0\cite{Mathematica}
and are provided in an accompanying notebook available for download from \href{https://doi.org/10.5281/zenodo.4765100}{https://doi.org/10.5281/zenodo.4765100}.
Atomic units are used throughout.

%%%%%%%%%%%%%%%%%%%%%%%%%%%%%%%%%%%%%%%%%%%%%
\section{Formulating Fractional Spins}
%%%%%%%%%%%%%%%%%%%%%%%%%%%%%%%%%%%%%%%%%%%%%

In the high- or low-spin ($\ms= \pm \frac{1}{2}$) configurations, the exact (ground state) wave function of a
one-electron system is represented by a single occupied orbital $\psi_0(\br)$, where $\br$ is the 
electronic position vector.
The corresponding spatial electron density is then
\begin{equation}
\rho_0(\br) = \abs{\psi_0(\br)}^2, 
\end{equation}
with the normalization constraint $\int \rho_0(\br)\,\dd \br = 1$.
The corresponding exact energy is given by the kinetic $T[n_0]$ and external potential $\vext[\rho_0]$ terms as
\begin{equation}
E_0 = \ke[\rho_0] + \vext[\rho_0] \equiv h[\rho_0],
\end{equation}
where we introduce the combined one-body component of the energy as $h[\rho_0]$.

At fractional-spin values ($ -\frac{1}{2} < \ms < \frac{1}{2}$), the electron density can be represented by 
a two-state weighted ensemble of a high-spin and low-spin density as
\begin{equation}
\rhow(\br) = (1-w)\, \rhoa(\br) + w\, \rhob(\br).
\label{eq:rhoUHF}
\end{equation} 
with $0 \le w \le 1$.
In the spin-unrestricted framework, the high-spin and low-spin densities are built from different orbitals $\psia(\br)$ and $\psib(\br)$ as
$\rhos(\br) = \abs{\psis(\br)}^2$ (with $\sigma = \alpha$ or $\beta$), providing individually normalized densities, \ie, $\int \rhos(\br)\,\dd \br = 1$.
The ensemble density is therefore normalized, \ie, $\int \rhow(\br)\,\dd \br = 1$, and  
spin-pure states are recovered at $w = 0$ and $w = 1$, with $\ms= \pm \frac{1}{2}$.
The equiweight ensemble (\ie, $w = \frac{1}{2}$) is of particular interest as it corresponds to a closed-shell 
system containing half a spin-up and half a spin-down electron.
For any other weight, there is no spin symmetry. 
The {fractional-spin} error is maximum in the case of the equiweight ensemble,\cite{Cohen2008,Cohen2008b,Cohen2012} as we shall illustrate below.

The simplest way to build the high-spin and low-spin spatial densities is to introduce 
the constraint $\rhoa(\br) = \rhob(\br) \equiv \rho(\br)$ and build the spatial density from a single occupied orbital
$\psi(\br)$.
This approximation leads to a spin-restricted framework with the corresponding RHF ensemble energy
\begin{equation}
\begin{split}
E_\RHF^w &= h[\rho] + (1-w)w\, (\psi\psi|\psi\psi).
\end{split}
\label{eq:RHFenergy}
\end{equation}
Here, the two-electron integrals are defined generally as
\begin{equation}
(\psi_p\psi_q|\psi_r\psi_s) = \iint \frac{\psi_p^*(\br_1) \psi_q^{\vphantom{*}}(\br_1) \psi_r^*(\br_2) \psi_{s}^{\vphantom{*}}(\br_2)}{r_{12}} \dd \br_1 \dd \br_2,
\end{equation}
with the inter-electronic distance $r_{12} = \abs{\br_1 - \br_2}$.
Equation~\eqref{eq:RHFenergy} shows that the exact energy is only recovered 
for $w = 0$ or 1, while an artificial {Coulomb interaction} arises for all other $w$ values.
This {fractional-spin} error occurs because the exchange interaction between the high- 
and low-spin components of the electron density (which in this case represent the same electron) is missing from 
the restricted ensemble energy. 
As a result, the artificial Coulomb repulsion is not completely cancelled.

Alternatively, allowing the high-spin and low-spin densities to differ leads to the UHF ensemble energy
\begin{equation} 
E_\UHF^w 
= (1-w)h[\rhoa] + wh[\rhob] 
+ (1-w)w(\psia \psia | \psib \psib).
\label{eq:UHFenergy}
\end{equation}
In Fukutome's classification, the real UHF solution is described as an ``axial spin density wave''. \cite{Fukutome1981}
From Eq.~\eqref{eq:UHFenergy}, we see that the {fractional-spin} error for $0<w<1$  
corresponds to an artificial Coulomb repulsion between the high-spin and low-spin 
components of the density.
In contrast to the restricted ensemble, the unrestricted orbitals allow this error
to be reduced through independent spatial relaxation of the high- and low-spin electron densities.
As we shall see later, the extent to which this relaxation can eradicate 
the {fractional-spin} error at a fixed value of $w$ is greatly increased when spatial separation of the high- and low-spin densities becomes possible. 
Finally, using a GHF-based approach for the fractional-spin leads to the correct energy for all $w$
(see Appendix~\ref{apdx:GHF}).

The {artifical interaction} term in Eqs.~\eqref{eq:RHFenergy} and \eqref{eq:UHFenergy} is reminiscent of the ghost-interaction in ensemble DFT that
causes the non-linearity of the ensemble energy for approximate density functionals. \cite{Gidopoulos2002,Pastorczak2014,Alam2016,Alam2017,Gould2017,Loos2020,Marut2020} 
Under the exact exchange-correlation functional, the fractional-spin ensemble density must yield the exact energy
for all $w$.\cite{Perdew1982,Cohen2008}
As in Gross--Oliveira--Kohn DFT (GOK-DFT) \cite{Gross1988a,Gross1988b,Oliveira1988} and $N$-centered ensemble DFT \cite{Senjean2018,Senjean2019} [and unlike in the original Perdew--Parr--Levy--Balduz (PPLB) theory \cite{Perdew1982}], the exact exchange-correlation functional is therefore weight-dependent as the total density (which integrates to the same number of electrons for all weights) does not contain enough information to unambiguously describe all the states belonging to the ensemble.
However, this constancy condition of the ensemble energy (which is a particular case of the piecewise linearity of the ensemble energy \cite{Perdew1982,Cohen2008c,Kraisler2013,Kraisler2014}) is not fulfilled for any currently available approximate functionals, \cite{Cohen2007,Kraisler2015a,Kraisler2015b} including HF theory. 
The resulting energy deviation is usually equated with a static correlation error.\cite{Cohen2008,Cohen2008b}

%%%%%%%%%%%%%%%%%%%%%%%%%%%%%%%%%%%%%%%%%%%%%
\section{Hydrogen Atom}
\label{sec:Hatom}
%%%%%%%%%%%%%%%%%%%%%%%%%%%%%%%%%%%%%%%%%%%%%

%==========================
\subsection{Ground State Energy}
\label{sec:HatomGS}
%==========================

%%%%%%%%%%%%%%%%
% RHF vs UHF fractional spin
%%%%%%%%%%%%%%%%
\begin{figure*}[ht]
\includegraphics[width=0.65\linewidth]{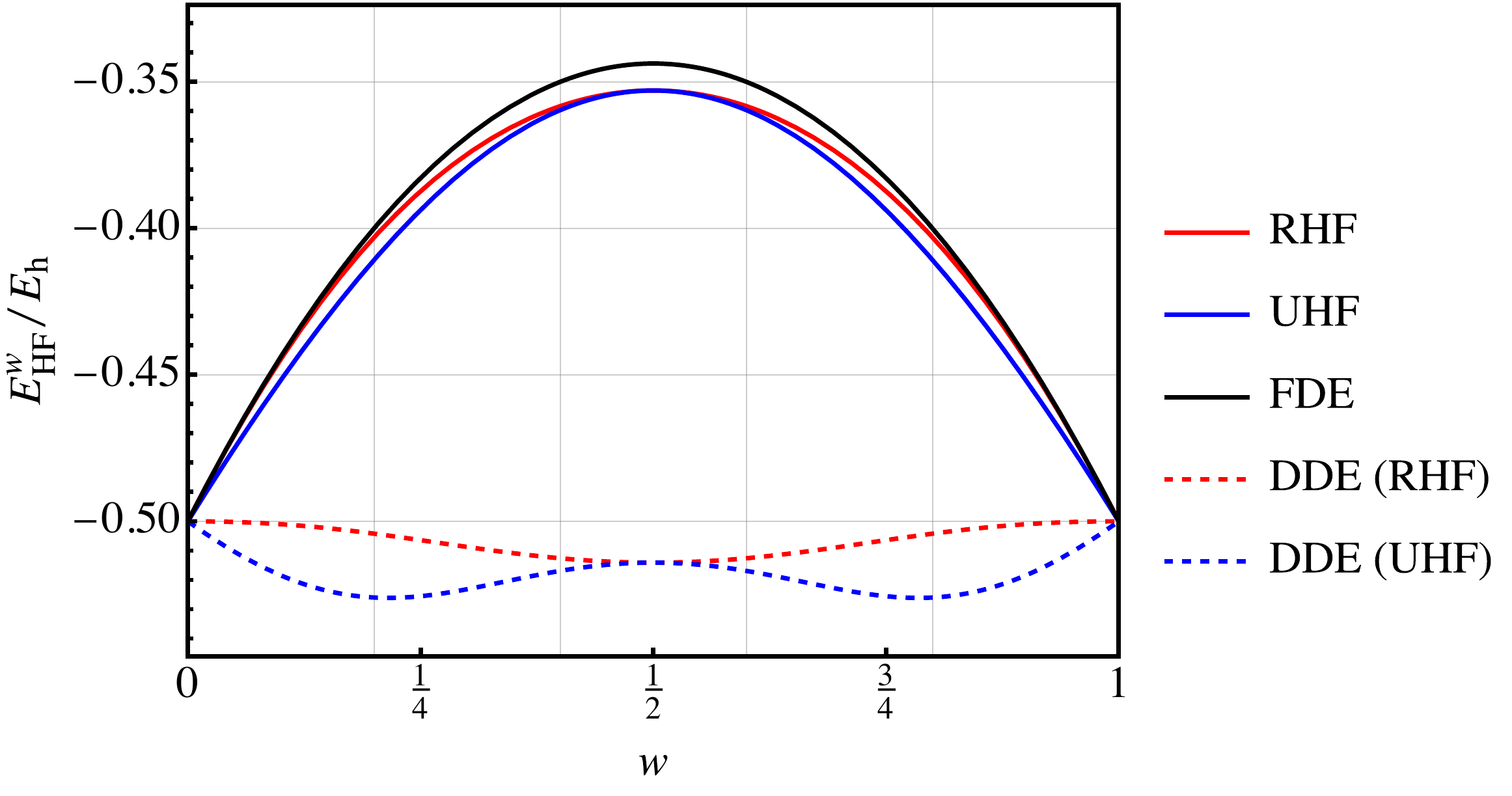}
\caption{Restricted (solid red) and unrestricted (solid blue) ensemble energies for the fractional-spin \ce{H} atom at the Hartree--Fock level. 
The functional-driven error (FDE) and the RHF and UHF density-driven errors (DDE) are also represented (see main text for more details).}
\label{fig:rhf_uhf}
\end{figure*}
%%%%%%%%%%%%%%%%

To illustrate the fractional-spin error for different HF formalisms, 
we first consider the hydrogen atom in a minimal spatial basis set comprising the lowest-energy 
$1\s$ and $2\s$ atomic orbitals
\begin{subequations}
\begin{align}
\chi_{1\s}(\br) &= \sqrt{\frac{1}{4\pi}} \exp(-r),
\\
\chi_{2\s}(\br) &= \sqrt{\frac{1}{32\pi}} (2-r)\exp(-r/2).
\end{align} 
\end{subequations}
Here, $r = \abs{\br}$ defines the scalar electron--nucleus distance.
The same physics occurs in larger basis sets and some results obtained in a large even-tempered basis \cite{Schmidt1979} are reported in the \SupInt.

In this minimal basis, the single occupied orbital for the fractional-spin RHF ensemble can be parametrized as 
\begin{equation}
\psi(\br) = \cos \theta \, \chi_{1\s}(\br) + \sin \theta \, \chi_{2\s}(\br),
\end{equation}
where $\theta$ represents an orbital rotation angle to be optimized.
Alternatively, the high-spin and low-spin orbitals in an unrestricted ensemble may be built using
different rotation angles for each spin, $\ta$ and $\tb$, to give
\begin{subequations}
\label{eq:psiUHF}
\begin{align}
\psia(\br) &= \cos \ta \, \chi_{1\s}(\br) + \sin \ta \, \chi_{2\s}(\br),
\\
\psib(\br) &= \cos \tb \, \chi_{1\s}(\br) + \sin \tb \, \chi_{2\s}(\br).
\end{align}
\end{subequations}

% RHF Energetic Error
In Fig.~\ref{fig:rhf_uhf}, we compare the optimized energies for the restricted (solid red) and
unrestricted (solid blue) fractional-spin ensembles.
At the spin-pure values  $w = 0$ or $1$, both representations are equivalent and yield the exact
hydrogen ground-state energy $E_0 = \SI{-1/2}{\hartree}$.
However, at fractional-spin values with $0 < w < 1$, we find the previously observed 
fractional-spin error,\cite{Cohen2008} reaching a maximum at $w=\frac{1}{2}$.
The restricted and unrestricted ensembles yield equivalent results for the spin-pure states 
at $w=0$ and $1$ ($\ms = \pm\frac{1}{2}$) and the spin-unpolarized state at $w = \frac{1}{2}$ ($\ms=0$).
However, the greater flexibility of the unrestricted ensemble leads to some energetic relaxation for intermediate $w$ values,
although this effect is marginal in comparison to the total fractional-spin error.

% RHF Density Relaxation
Under the exact functional, or within the GHF formalism (see Appendix~\ref{apdx:GHF}), the spatial density $\rho_0(\br)$ remains the same for all $w$.
Fixing the spatial density at its exact value $\rho(\br) = \rho_0(\br)$ and computing the energy \textit{via} the (incorrect) HF energy functional
allows the functional-driven error (FDE) to be defined, with the remaining part of the error being the density-driven error (DDE).\cite{Kim2013,Vuckovic2019} In Ref.~\onlinecite{Vuckovic2019} the relation between DDE and static correlation has been analyzed for RHF using the Hubbard dimer, where the DDE has been shown to be very small and becomes zero for the symmetric dimer. In Fig.~\ref{fig:rhf_uhf}, we find that the fractional-spin DDE is also very small in the
\ce{H} atom, and thus this system is dominated by the FDE. 
However, in contrast to the symmetric Hubbard dimer, the DDE in RHF for the \ce{H} atom is not zero and becomes maximally large at $w=\frac{1}{2}$.
Furthermore, the DDE in UHF is always larger (in magnitude) than RHF, reaching maximum magnitude
when the UHF energetic stabilization is most significant.
In other words, the energetic stabilization provided by the unrestricted approach has a detrimental effect on the quality of the electron density,\cite{Medvedev2017}
while the restricted ensemble has a greater energy error but a more accurate density.
This result is consistent with the broader symmetry-dilemma in HF theory, where lower electronic energies 
are achieved at the expense of less accurate electronic densities and the loss of well-defined quantum numbers.\cite{Lykos1963}

%%%%%%%%%%%%%%%%%%%%%%%%%%%
\subsection{Unrestricted Density Relaxation}
%%%%%%%%%%%%%%%%%%%%%%%%%%%

%%%%%%%%%%%%%%%%%%%%%%%%%%%
\begin{figure}[b]
\includegraphics[width=\linewidth]{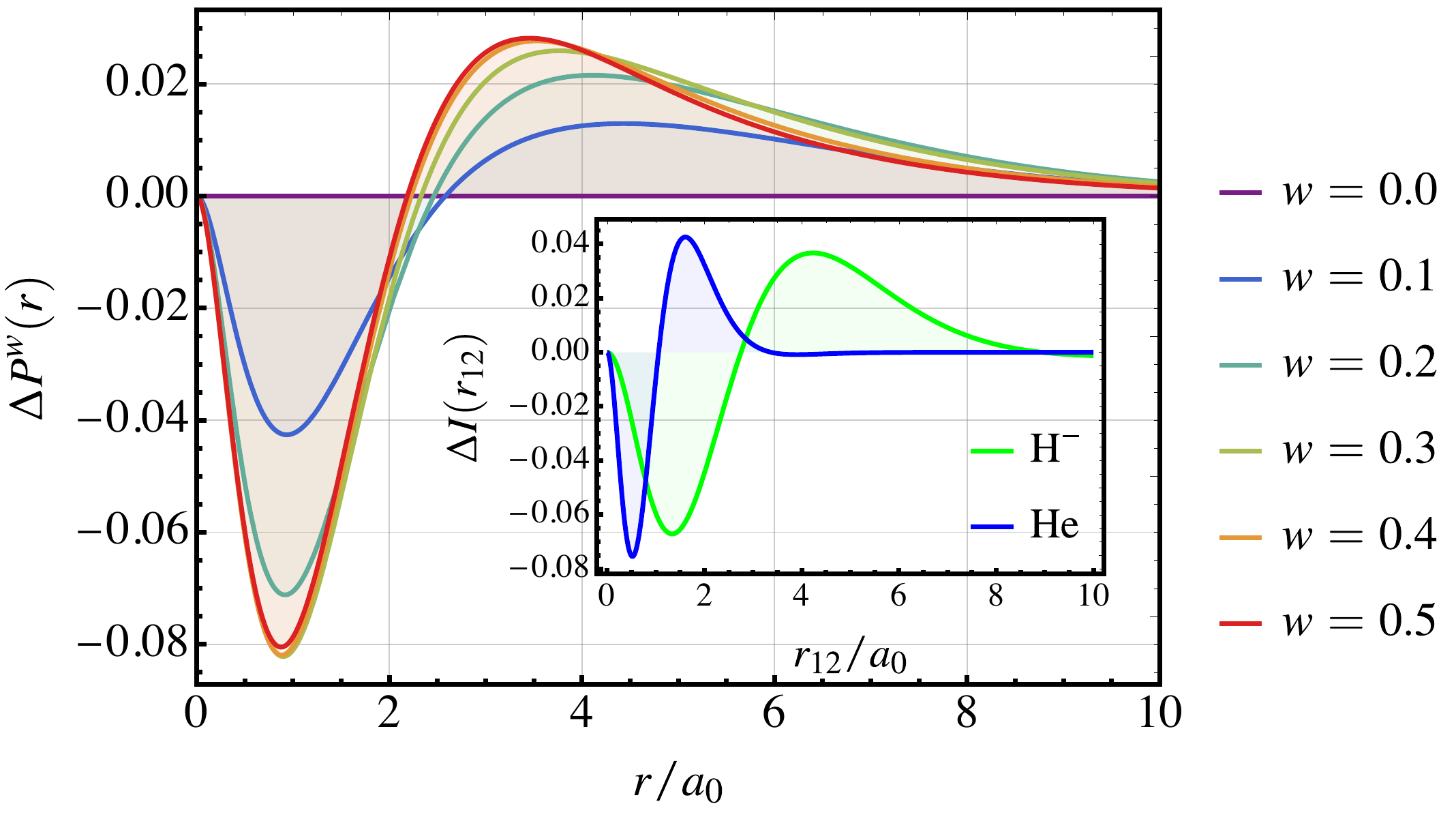}
\caption{Difference between unrestricted and exact radial probability densities for various weights $w$.
The density depletion for $w > 0$ is reminiscent of the system-averaged Coulomb hole in \ce{He}-like ions shown in the inset for He and H$^-$, where we report the difference between accurate correlated intracule densities and the corresponding Kohn--Sham ones (both taken from Ref.~\onlinecite{Gori-Giorgi2005}).}
\label{fig:rhf_relax}
\end{figure}
%%%%%%%%%%%%%%%%%%%%%%%%%%%

In Fig.~\ref{fig:rhf_relax}, we explore the UHF density relaxation by computing the change 
in the radial probability density for different $w$ values, defined as 
\begin{equation}
\label{eq:Pw}
\Delta P^w(r) = 4\pi\,r^2\,\qty[\rhow(r) -  \rho_0(r)].
\end{equation}
As $w$ increases from $0$ to $\frac{1}{2}$, the fractional-spin electrons shift away from the nucleus to 
become less dense and reduce the strength of the artificial Coulomb repulsion.
While this relaxation decreases the {fractional-spin} error, 
it also reduces the strength of the electron-nuclear attraction. 
If the nuclear charge $Z$ is increased continuously to non-integer values, it becomes increasingly less favorable for the electrons to
relax away from the nucleus and optimizing the density does not reduce the {fractional-spin} error 
as much. 
On the contrary, if the nuclear charge is reduced to $Z < 1$, it becomes more 
energetically favorable for the electron density to relax away from the nucleus.
Eventually, relaxation becomes so favorable that the fractional-spin electrons are ionized from 
the nucleus. 
This analysis explains the existence of an artificial critical nuclear charge for the bound-state stability of the 
one-electron atom that has recently been observed using numerical fractional-spin RHF 
calculations at $w = \frac{1}{2}$.\cite{Burton2021b}

Figure \ref{fig:rhf_relax} also illustrates a close similarity between the difference of density probabilities in the unpolarized \ce{H} atom and the system-averaged Coulomb hole 
in the \ce{He}-like ions,\cite{Coulson1961}
defined as the difference between the exact intracule $I(r_{12})$ and either the HF or the Kohn--Sham intracule $I_0(r_{12})$ as
\begin{equation}
\Delta I(r_{12})= I(r_{12}) -  I_0(r_{12}).
\end{equation}
The intracule gives the probability of finding two electrons at a distance $r_{12}$ and has been extensively used to study the physics of electronic correlation. \cite{Gill2006,Gori-Giorgi2008,Pearson2009,Per2009,Via-Nadal2019,Rodriguez-Mayorga2019}
In \ce{He}-like ions, the Coulomb hole results from the electrons being too close together at the mean-field HF or Kohn--Sham level. 
Electron correlation pushes the electrons apart, creating a depletion of probability density at short inter-electronic distances.
The same physics are clearly involved in the fractional-spin H atom: 
the ensemble HF approximation pushes the electrons away from each other to lower their {artifical Coulomb interaction}.
In this sense, the fractional-spin one-electron atom behaves as a two-electron system with electron  correlation. 

%%%%%%%%%%%%%%%%%%%%%%
\begin{figure}[t]
\includegraphics[width=\linewidth]{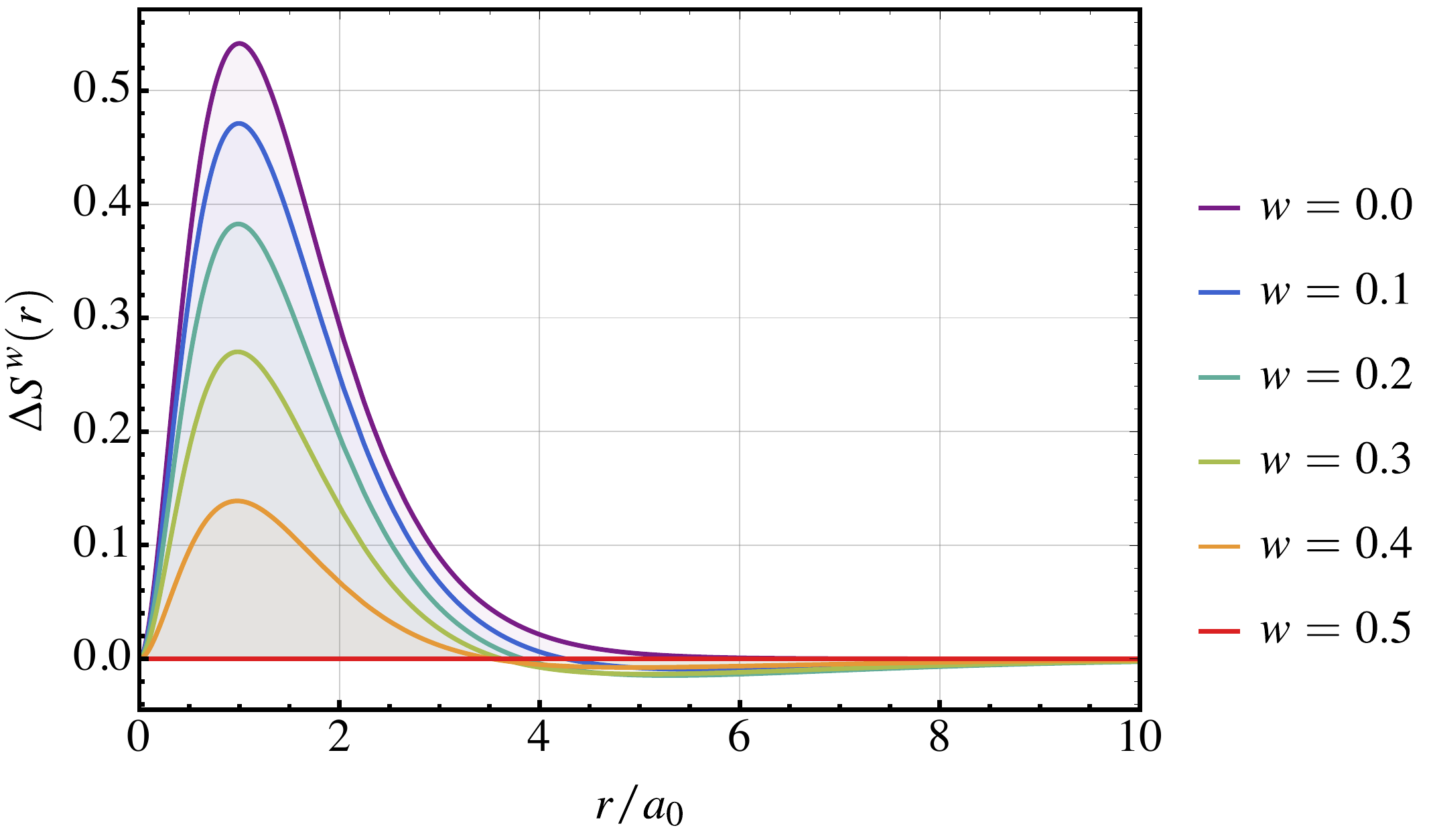}
\caption{Radial probability spin density of the unrestricted ground-state fractional-spin ensemble 
in the \ce{H} atom for various $w$.
For $w<0.5$, the dominant high-spin density localizes closer to the nucleus to maximize the electron-nuclear attraction.}
\label{fig:uhf_relax}
\end{figure}
%%%%%%%%%%%%%%%%%%%%%%

%%%%%%%%%%%%%%%%
\begin{figure*}
\includegraphics[width=\linewidth]{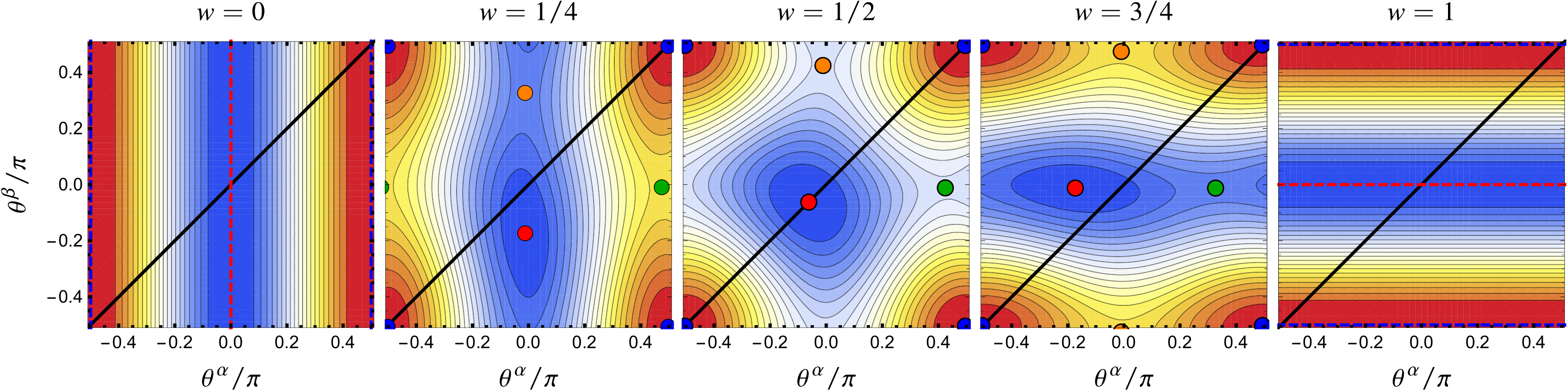}
\caption{Unrestricted ensemble energy surface for the fractional-spin \ce{H} atom 
as a function of the orbital rotation angles $\ta$ and $\tb$.
The restricted ensembles are indicated by the black line $\ta = \tb$.
At $w=0$ and $1$, the energy is invariant to $\tb$ and $\ta$ respectively, and stationary points occur
as continuous ``ridge'' (red dashed) lines.
For intermediate values, four discrete stationary points occur (colored dots) with additional periodic repeats.
Here, the different colors of stationary points correspond to the multiple solutions shown in Fig.~\ref{fig:excited_uhf}.
}
\label{fig:uhf_surf}
\end{figure*}
%%%%%%%%%%%%%%%%

In Fig.~\ref{fig:uhf_relax}, we consider the radial spin-density probability for different $w$ values, defined as
\begin{equation}
\Delta S^w(r) = 4\pi\,r^2\, \qty[(1-w)\,\rhoa(r) - w \, \rhob(r)] .
\end{equation}
Although it is not explicitly specified, both $\rhoa(r)$ and $\rhob(r)$ depend implicitly on $w$ as the underlying orbitals $\psia(r)$ and $\psib(r)$ are obtained with a weight-dependent energy functional.
For $w < \frac{1}{2}$, the majority high-spin density preferentially occupies the region closest to the nucleus with 
the minority low-spin density at larger radial distances.
This spatial separation of the high- and low-spin density minimizes the (artificial) Coulomb repulsion
between the two spin components while maximizing the overall electron-nuclear attraction.
The magnitude of this spin-polarization decreases as $w$ increases until eventually yielding a spin-unpolarized (restricted) density at $w = \frac{1}{2}$.
Spatial separation of the different spin components is therefore essential for providing the additional relaxation of the unrestricted ensemble relative to the restricted ensemble.

% STRANGE SYMMETRY BREAKING
The additional relaxation of the unrestricted ensemble appears analogous to symmetry-broken wave
functions in strongly correlated molecules, such as the dissociation limit of \ce{H2}.
In \ce{H2}, a lower-energy UHF solution emerges at the Coulson--Fischer point to become the global HF ground-state,\cite{SzaboBook,Coulson1949}
with the high-spin and low-spin electrons localized on opposite atoms, while the delocalized RHF solution becomes a saddle point of the UHF energy.
However, in the fractional-spin \ce{H} atom, the restricted ensemble
is only a stationary point of the UHF energy at $w = 0$, $\frac{1}{2}$ and $1$.
This can be seen by considering the electronic energy landscape\cite{Burton2021} of the unrestricted ensemble as a function of $\ta$ and $\tb$, 
as shown in Fig.~\ref{fig:uhf_surf}.
For all other values of $w$, the stationary points on the unrestricted energy surface 
do not correspond to restricted ensembles with $\ta = \tb$ (black line).
Consequently, the optimal restricted ensemble is only a constrained minimum along the line
$\ta = \tb$, rather than a saddle point of the UHF energy.
This unusual type of instability in HF is reminiscent of open-shell systems where the restricted open-shell HF (ROHF)
is not a stationary point of the UHF energy,\cite{Andrews1991} and is likely to be the case in many open-shell systems.
Alongside the ground-state ensemble, there are additional stationary points on the unrestricted ensemble energy that 
we will address in Sec.~\ref{sec:HatomES}.

%==========================
\subsection{Spin Components}
\label{sec:spinH}
%==========================

With different densities for different spins, the restricted and unrestricted ensembles both 
have the potential for introducing spin-contamination.
Analytic formulae for the expectation values of the square of the spin operator $\expval*{\cS^2} = \expval*{\cS_x^2 + \cS_y^2 + \cS_z^2}$ can be derived (see Appendix~\ref{apdx:S2equations}) for the RHF and UHF ensembles as 
\begin{subequations}
\begin{align}
\expval*{\cS^2}_\UHF & = \expval*{\cS^2}_\text{exact} - \frac{1}{2} w(1-w)\qty(1+ 2\abs{S_{\alpha \beta}}^2),
\label{eq:S2_UHF}
\\
\expval*{\cS^2}_\RHF & = \expval*{\cS^2}_\text{exact} - \frac{3}{2}w(1-w),
\label{eq:S2_RHF}
\end{align} 
\end{subequations} 
where $S_{\alpha \beta} = \braket*{\psi_\alpha}{\psi_\beta}$ is the spatial overlap between the spin-up and spin-down orbitals, and $\expval*{\cS^2}_\text{exact} = \frac{3}{4}$.
The corresponding GHF expectation value $\expval*{\cS^2}_\GHF$ is exact and independent of $w$, with the individual components
\begin{equation}
\expval*{\cS_x^2}_\text{exact} = \expval*{\cS_y^2}_\text{exact} = \expval*{\cS_z^2}_\text{exact} = \frac{1}{4}.
\end{equation}
In contrast, the individual components in the restricted and unrestricted approximations are given by
\begin{subequations}
\begin{align}
\expval*{\cSx^2}_\UHF &= \expval*{\cSx^2}_{\text{exact}} - \frac{1}{2}w(1-w)\abs{S_{\alpha \beta}}^2,
\label{eq:Sx2_UHF}
\\
\expval*{\cSy^2}_{\UHF} &= \expval*{\cSy^2}_{\text{exact}} - \frac{1}{2}w(1-w)\abs{S_{\alpha \beta}}^2,
\\
\expval*{\cSz^2}_\UHF &= \expval*{\cSz^2}_{\text{exact}} -\frac{1}{2}w(1-w),
\label{eq:Sz2_UHF}
\end{align} 
\end{subequations} 
and
\begin{align}
\expval*{\cS_{\tau}^2}_\RHF &= \expval*{\cS_{\tau}^2}_\text{exact} - \frac{1}{2}w(1-w),
\label{eq:Sx2_RHF}
\end{align} 
where $\tau \in \{x,y,z\}$.

%%%%%%%%%%%%%%%%%%%%%%
\begin{figure}[b]
\includegraphics[width=\linewidth]{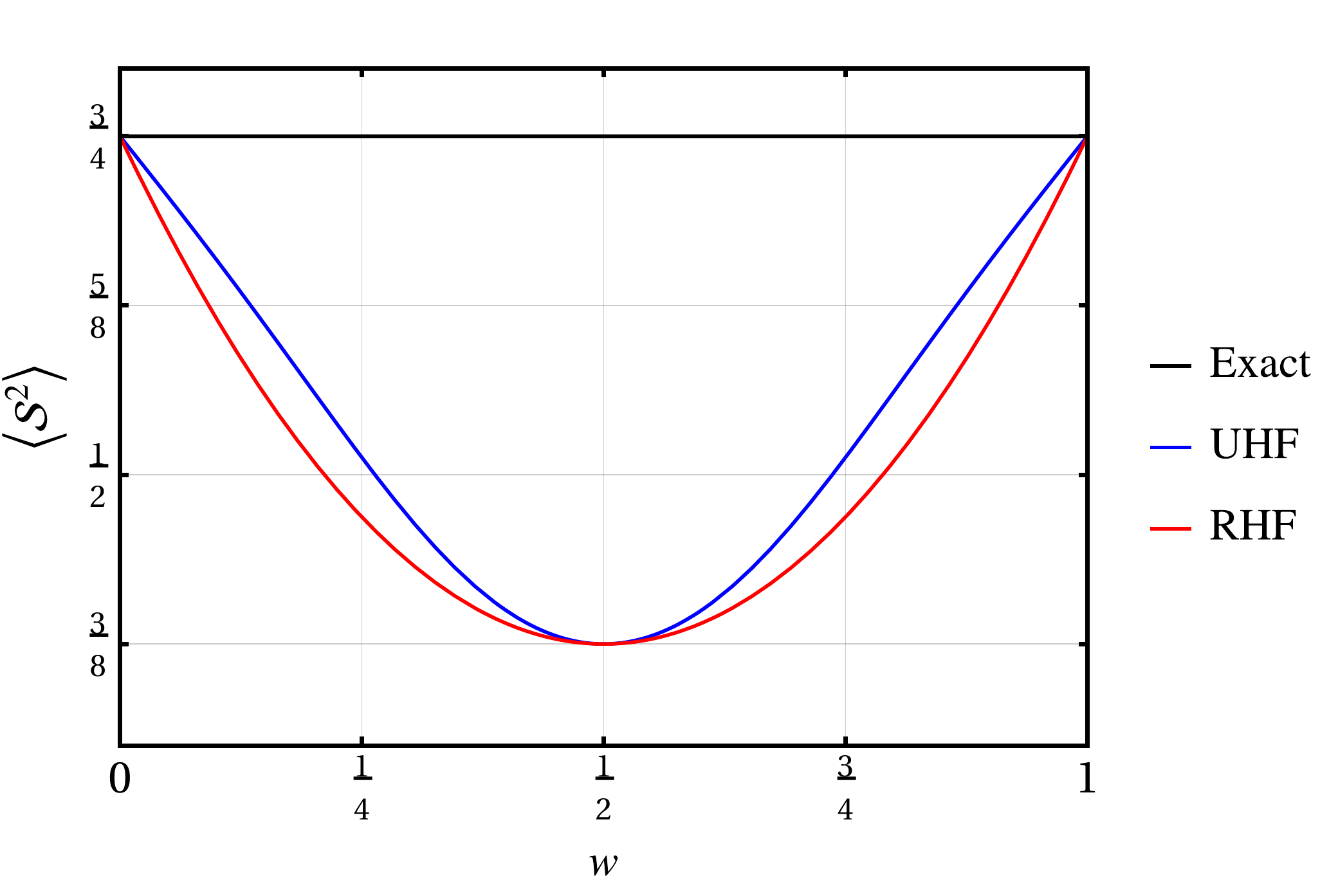}
\caption{Spin expectation values in the \ce{H} atom as functions of the weight $w$.
Both the restricted and unrestricted ensemble approximations result in spin contamination.}
\label{fig:spin}
\end{figure}
%%%%%%%%%%%%%%%%%%%%%%

For the restricted ensemble, the spin contamination adopts a quadratic form that reaches $\expval*{\cS^2} = \frac{3}{8}$ at $w = \frac{1}{2}$, which is exactly half the physically correct value (see Fig.~\ref{fig:spin}).
We believe that the $w=\frac{1}{2}$ ensemble can be interpreted as an equally-weighted sum of the exact singlet states corresponding to the \ce{H+} nucleus with no electrons $n^{\boldsymbol{\cdot} \boldsymbol{\cdot}} $ and the two-electron  \ce{H-} ground state $n^{\uparrow \downarrow}$, and the one-electron doublet states corresponding to either a spin-up electron $n^{\uparrow \boldsymbol{\cdot}}$ or spin-down electron $ n^{\boldsymbol{\cdot} \downarrow}$, giving
\begin{equation}
\rhow = \frac{1}{4} \qty(n^{\boldsymbol{\cdot} \boldsymbol{\cdot}} + n^{\uparrow \boldsymbol{\cdot}} + n^{\boldsymbol{\cdot} \downarrow}  + n^{\uparrow \downarrow}).
\end{equation}
Taking the weighted sum of expectation values then yields
\begin{equation}
\expval*{\cS^2} = \frac{1}{4}\qty( 0 + \frac{3}{4} + \frac{3}{4} + 0 ) = \frac{3}{8}.
\end{equation}
This result can be compared with the symmetry-broken UHF wave function in the dissociation limit of \ce{H2} where the value $\expval*{\cS^2} = 1$ indicates that the 
wave function is an equal combination of a singlet and triplet state.\cite{SzaboBook}
However, when fractional numbers of electrons are allowed, it appears that the ensemble can become a combination of exact densities with different numbers of particles.
A detailed study of this intriguing feature of fractional spins is left for future work.

With the unrestricted orbitals, spatial separation generally gives $\abs{S_{\alpha \beta}}^2 < 1$, hence reducing
the spin-contamination in the $\expval*{\cSx^2}$ and $\expval*{\cSy^2}$ components.
This result is surprising and counter-intuitive as, elsewhere, spin-contamination is generally found to \emph{increase} in an unrestricted representation. \cite{Krylov2000,Lee2018,Casanova2020}
Furthermore, using $\expval*{\cSz}_{\UHF} = \expval*{\cSz}_\RHF = \frac{1}{2} - w$, the variance of $\cSz$ is given in both the restricted and unrestricted approximations as 
\begin{equation}
\expval*{\cSz^2} - \expval*{\cSz}^2 = \frac{1}{2}w(1-w),
\end{equation}
confirming that the ensemble densities are only eigenstates of $\cSz$ at $w=0$ or $1$.

%%%%%%%%%%%%%%%%%%%%%%
\begin{figure}[t]
\includegraphics[width=0.8\linewidth]{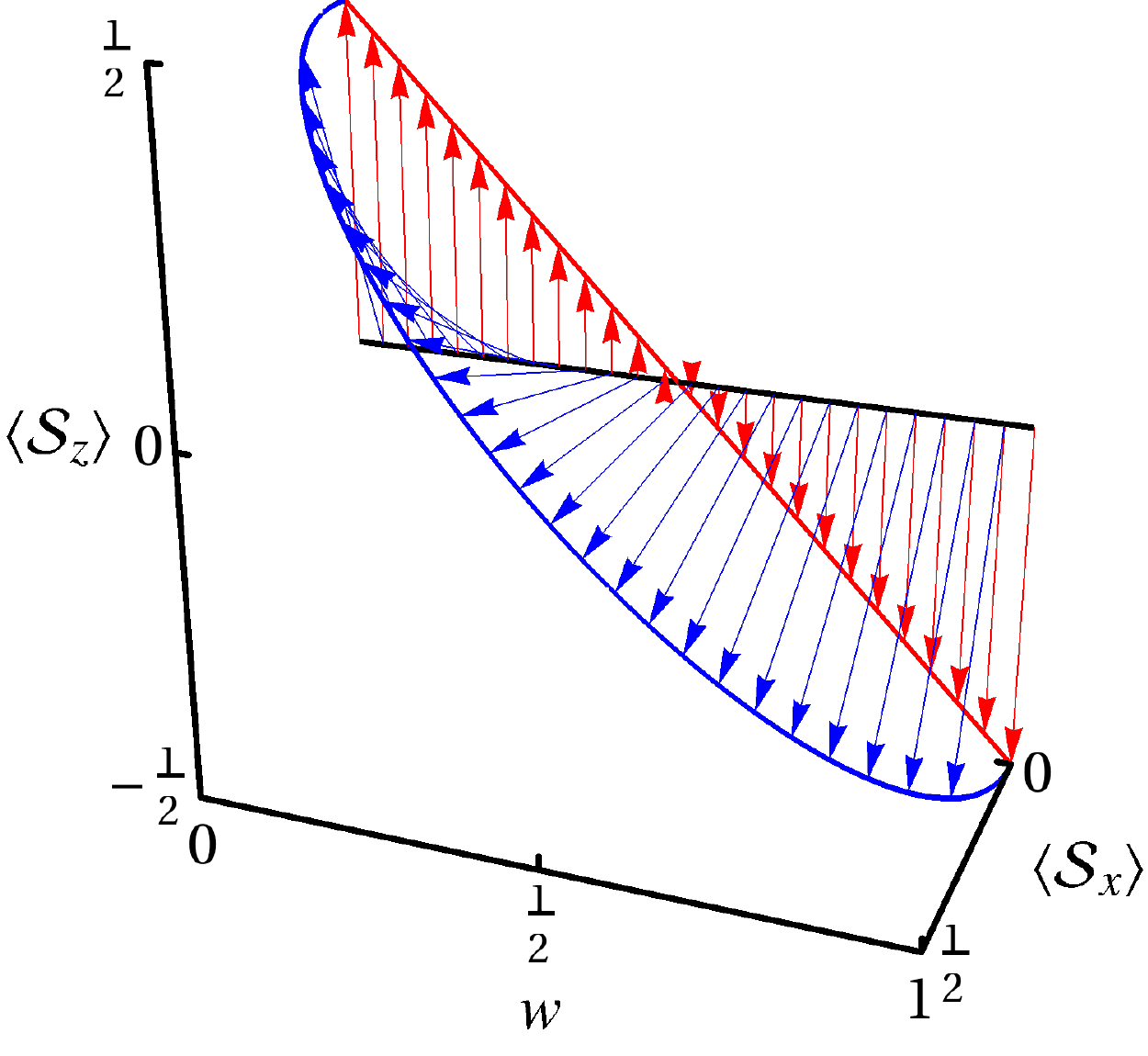}
\caption{$\expval*{\cS_z}$ and $\expval*{\cS_x}$ as functions of the weight $w$ for RHF/UHF (red) and GHF (blue).}
\label{fig:corkscrew}
\end{figure}
%%%%%%%%%%%%%%%%%%%%%%

The spin-contamination in the RHF and UHF approximations has exactly the same origin as the {fractional-spin} error in the 
energy: an ``exchange'' contribution to $\expval*{\cS^2}$ that corresponds to the spin coupling of the spin-up and spin-down components of the density is missing.
The spin expectation values can therefore be used to understand how GHF is able to provide the correct exchange energy to fully 
cancel the {fractional-spin} error.
For this, we turn to the individual spin components.	 
Figure \ref{fig:corkscrew} shows $\expval*{\cS_z}$ and $\expval*{\cS_x}$ as functions of $w$ for the RHF/UHF and GHF approximations.
In the RHF/UHF case (red curve in Fig.~\ref{fig:corkscrew}), the (fractionally-occupied) orbitals are constrained to be eigenfunctions of the $\cS_z$ operator and we find $\expval*{\cS_x} = \expval*{\cS_y} = 0$ and $\expval*{\cS_z} = \frac{1}{2}-w$.
The overall spin-vector therefore remains (anti-)parallel to the $z$-axis throughout the transformation from $\expval*{\cS_z} = \frac{1}{2}$ to $-\frac{1}{2}$.
In contrast, the additional flexibility of the GHF approximation (blue curve in Fig.~\ref{fig:corkscrew}) allows the electron spin vector to rotate in the $xz$-plane with components $\expval*{\cS_x}_\text{GHF} = \sqrt{(1-w)w}$ and $\expval*{\cS_z}_\text{GHF} = \frac{1}{2} - w$ (blue curve in Fig.~\ref{fig:corkscrew}).
This rotation conserves the overall norm of the spin vector and results in an exchange interaction that cancels out the Coulomb interaction.
The curves in Fig.~\ref{fig:corkscrew} further illustrate Fukutome's classification of the UHF and GHF approximations as axial and torsional spin density waves, respectively.\cite{Fukutome1981}

%==========================
\subsection{Excited States}
\label{sec:HatomES}
%==========================

\begin{figure}[t]
\includegraphics[width=\linewidth]{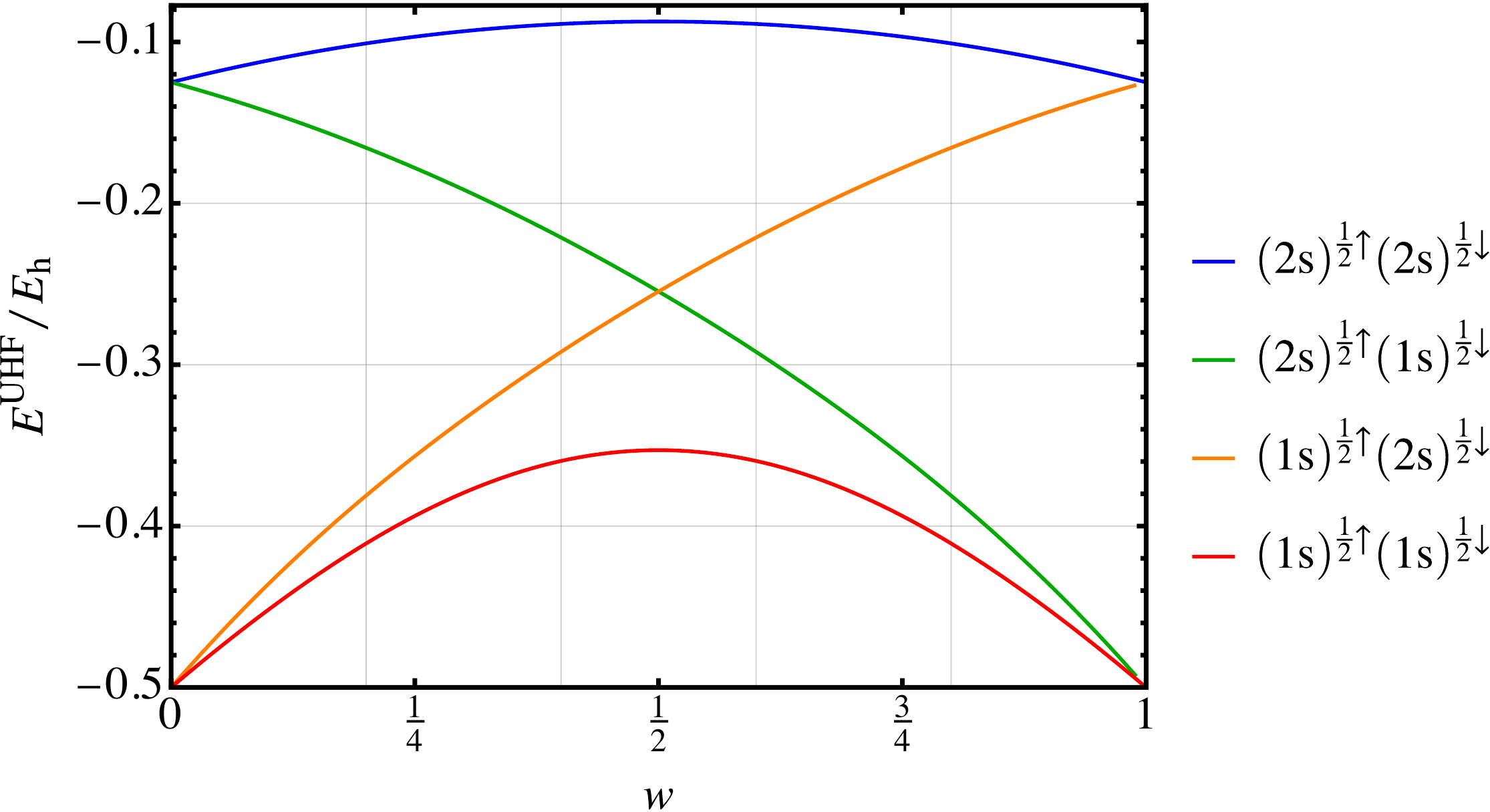}
\caption{Ensemble energy of the four UHF stationary points for the fractional-spin \ce{H} atom. 
Each solution corresponds to a stationary point of the UHF ensemble energy in Fig.~\ref{fig:uhf_surf}.
The electronic configuration of each unrestricted state for $w = \frac{1}{2}$ is given on the right-hand-side of the figure.}
\label{fig:excited_uhf}
\end{figure}

In Fig.~\ref{fig:uhf_surf}, we see that the lowest energy solutions
are not the only stationary points of the unrestricted ensemble energy surface. 
For $w = 0$ and 1, there is another solution with energy $\SI{-1/8}{\hartree}$ that corresponds to the physical 2s excited state.
At these limits, the energy becomes invariant to the orbital rotation angle representing the vacant spin orbital
($\tb$ for $w=0$ and $\ta$ for $w=1$) and  the $1\s$ and $2\s$ states occur as one-dimensional stationary ``ridges'' in the 
ensemble energy.
In contrast, for  $0 < w < 1$, we find that the $1\s$ and $2\s$ solutions become isolated stationary points 
on the energy surface, alongside a further two stationary points with intermediate energies, 
as shown in Fig.~\ref{fig:excited_uhf}.
Between $0< w< 1$, these additional solutions evolve smoothly from the $1\s$ ground state to the $2\s$ excited state 
(and vice versa) and can be qualitatively assigned to (fractional) open-shell $(1\s)^{\frac{1}{2}}(2\s)^{\frac{1}{2}}$ states at $w=\frac{1}{2}$.

The existence of these additional solutions is directly linked to the breakdown of the ensemble UHF equations for
fractional spins. 
While the optimized ground $1\s$ and excited $2\s$ solutions can be considered as an approximation to the corresponding exact
states, the $(1\s)^{\frac{1}{2}}(2\s)^{\frac{1}{2}}$-like solutions have no physical interpretation for a single electron.
We can therefore consider them as transient states that occur in regions where the unrestricted ensemble
is not exact.
The similarities between these configurations and the excited states of a two-electron atom 
further indicates that the fractional-spin HF ensemble behaves as a two-body problem.
Furthermore, these transient states create an interesting new connection between the physical ground and excited states.
Similar connecting pathways of stationary points can also be identified using non-Hermitian 
operators in the complex plane.\cite{Burton2019a,Marie2021}
In both contexts, these connections arise from an approximate description of electronic structure, and we predict that 
such phenomena will be present across a broad range of theoretical techniques.

%%%%%%%%%%%%%%%%%%%%%%
\subsection{M{\o}ller--Plesset Adiabatic Connection}
\label{sec:MPAC}
%%%%%%%%%%%%%%%%%%%%%%

Beyond the HF approximation, fractional-spin errors also play an important role in driving the physics of the restricted M{\o}ller--Plesset (MP) adiabatic connection, as discussed in Ref.~\onlinecite{Daas2020}.
In the MP adiabatic connection, a parametrized $\lambda$-dependent Hamiltonian is constructed to connect the zeroth-order MP reference states to the exact wave function at $\lambda~=~1$.
%{The conventional $n$th-order MP correction is then given by the derivatives 
%of the energy with respect to $\lambda$, evaluated at $\lambda = 0$.}
{The small-$\lambda$ series expansion for the eigenvalues of this Hamiltonian
corresponds to the conventional MP perturbation series, with the $n$th-order MP correction directly 
proportional to the $n$th derivative of the $\lambda$-parametrized energy at $\lambda=0$.}

{In Ref.~\onlinecite{Daas2020}, the MP adiabatic connection for the stretched \ce{H2} molecule is compared with twice the energy of the \ce{H} atom at $w=\frac{1}{2}$  (Fig.~10 of  Ref.~\onlinecite{Daas2020}), giving a perfect agreement as $R$ increases, except at $\lambda=0$ where the energy is discontinuous for $R\to\infty$. 
The conventional $n$th-order MP corrections %, corresponding to the energy derivatives with respect to $\lambda$,
are therefore not well-defined for $R \to \infty$, as expected because the restricted MP2 correlation energy 
%(which determines the slope of the curves in this figure) 
diverges to $-\infty$ in the \ce{H2} dissociation limit.}
In the current context, one particularly interesting result from Ref.~\onlinecite{Daas2020} is the suggestion that the spin of the $\lambda$-parametrized 
wave function in the fractional-spin \ce{H} atom will flip discontinuously as $\lambda$ passes through 1, {except for the special case $w=\frac{1}{2}$.}

While previously only the \ce{H} atom with $w=0$, $\frac{1}{2}$, and $1$ were considered,  here we extend this analysis to general $w$ values.
After finding the weight-dependent RHF spatial orbital $\psi$ (which is held fixed in the following equations), we consider the $\lambda$-dependent Hamiltonian
\begin{equation}\label{eq:HMPAC}
	\Hat{H}_\lambda^w=\hat{T}+\hat{V}_{\rm ext}+(1-\lambda)\qty(\hat{J}^w-\hat{K}^w),
\end{equation}
where the exact energy is recovered as the eigenvalue  at $\lambda = 1$.
Here, the multiplicative Hartree operator is defined as
\begin{equation}
	\Hat{J}^w=\int \frac{\abs{\psi(\br')}^2}{\abs{\br'-\br}} \dd \br',  
\end{equation}
and the non-local exchange operator $\hat{K}^w$ is defined by its action on a generalized orbital $\tpsi$ as
\begin{equation}
\begin{split}
\Hat{K}^w \tpsi= 
(1-w)\, \psi(\br) \ket{\alpha} 
&\int \frac{\psi^{*}(\br')\tpsi^{\alpha}(\br') }{\abs{\br'-\br}} \dd \br' 
\\
+ w\, \psi(\br) \ket{\beta}
&\int \frac{\psi^{*}(\br')\tpsi^{\beta}(\br') }{\abs{\br'-\br}} \dd \br'. 
\end{split}
\end{equation}
The generalized orbital $\tpsi$ can have both a high-spin ($\alpha$) and low-spin ($\beta$) component with
different spatial orbitals $\tpsi^{\alpha}$ and $\tpsi^{\beta}$, giving 
\begin{equation}
\tpsi(\bx) = \tpsi^{\alpha}(\br) \ket{\alpha} + \tpsi^{\beta}(\br) \ket{\beta},
\end{equation}
where $\bx$ represents the combined spin and spatial coordinates (see also Appendix \ref{apdx:GHF}).

In Ref.~\onlinecite{Daas2020}, the ground-state wave function $\psil$ and energy $E_\lambda^w$ were computed and analyzed 
in the range $\lambda\in [0,\infty)$ for $w=0, 1$ and $\frac{1}{2}$, while the spin of $\psil$ was constrained to be the same as the RHF orbital, essentially forbidding any spin flip. 
In this work, we release this constraint and give $\psil$ full variational freedom by defining
\begin{equation}\label{eq:psilambdaspin}
	\psil(\bx) = \psil(\br)\qty(\sqrt{1-q}\,|\alpha\rangle+\sqrt{q}\,|\beta\rangle)
\end{equation}
where $0 \le q \le 1$.
We can then analyze how the spin of $\psil$ changes with $w$ and $\lambda$. 
Note that, in the general many-electron case, the exact  $\psil$ is a correlated wave function
and Eq.~\eqref{eq:psilambdaspin} cannot be interpreted as a GHF state.
However, in this case, the two {are equivalent} because GHF is exact for one-electron systems.

%%%%%%%%%%%%%%%%%%%%%
\begin{figure*}[t]
	\includegraphics[width=\linewidth]{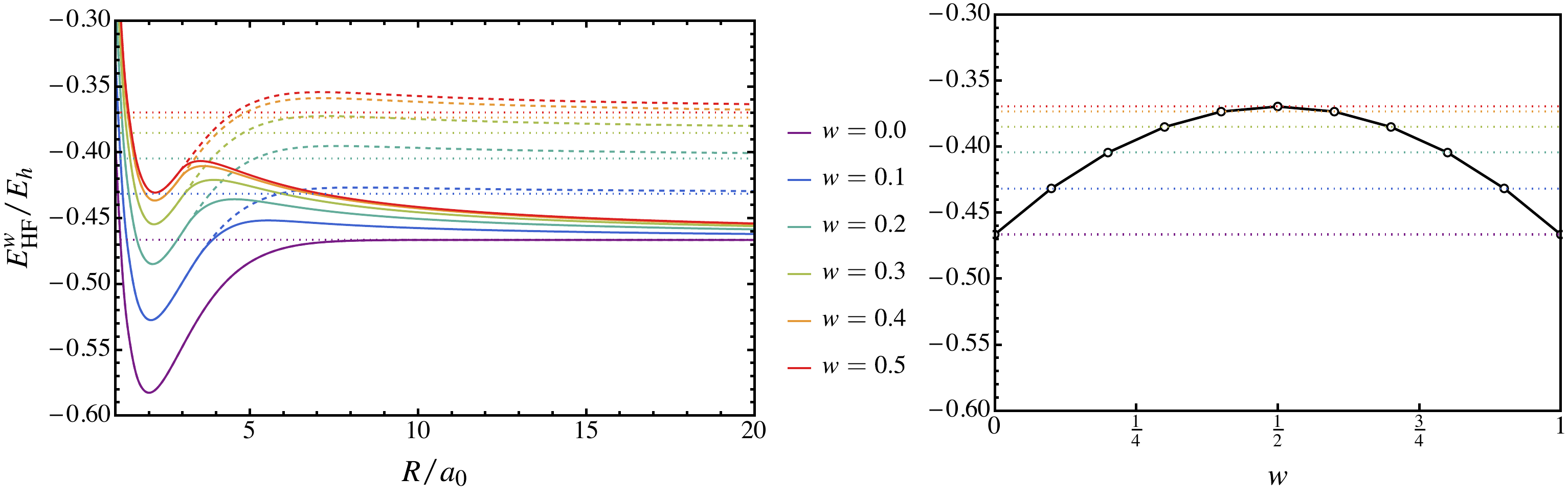}
\caption{Left: UHF (solid) and RHF (dashed) energies of \ce{H2+} as functions of the internuclear distance $R$ for various weights $w$. 
Right: RHF energies of \ce{H2+} in the dissociation limit as a function of the weight $w$.
Both graphs are computed with the STO-3G basis set.}
\label{fig:h2+}
\end{figure*}
%%%%%%%%%%%%%%%%%%%%%

Within the $w$- and $\lambda$-dependent Hamiltonian \eqref{eq:HMPAC}, only the expectation value of $\hat{K}^w_\lambda$ depends on $q$
\begin{equation}
	\mel{\psil}{\hat{K}^w}{\psil}=\qty[(1-q)(1-w)+q\,w](\psil \psi|\psil \psi).
\end{equation}
Since $\hat{K}^w$ is a positive-definite operator, we therefore find that the optimal value of $q$ depends on both $\lambda$ and $w$ as
\begin{subequations}
\begin{align}
{\rm for}\; \lambda < 1\text{:} \qquad & q =
\begin{cases}
1 & w > \frac{1}{2},
\\
0 & w < \frac{1}{2},
\end{cases}
\\
{\rm for}\; \lambda > 1\text{:} \qquad & q =
\begin{cases}
0 & w > \frac{1}{2},
\\
1 & w < \frac{1}{2}.
\end{cases}
\end{align}
\end{subequations}
In other words, the optimal $\psil$ is always in either a pure high-spin or low-spin configuration, and the spin will flip relative to the reference RHF orbital if 
$\lambda$ crosses 1 at fixed $w$ or if $w$ crosses $\frac{1}{2}$ at fixed $\lambda$.
At $\lambda=1$ and/or at $w=\frac{1}{2}$, the energy is independent of $q$. 
At $w=0$ or 1, the spin-flip that occurs when $\lambda$ crosses 1 is analogous to a similar phenomenon in the MP adiabatic connection 
for the stretched \ce{H2} molecule with a UHF reference, in which the two spins are exchanged between the two nuclei.\cite{Marie2021}
This feature is also responsible for the slow convergence of the unrestricted MP series in stretched \ce{H2}, as discussed for the Hubbard dimer in Ref.~\onlinecite{Marie2021},
and is intimately related to a critical point singularity in MP theory.\cite{Stillinger2000,Goodson2004,Sergeev2005,Sergeev2006}

Finally, we note that forbidding spin-flip as $\lambda$ changes is equivalent to setting $q=w$. 
Under this constraint, the prefactor for the expectation value of $\hat{K}^w$ becomes $s=1-2\,w+2\,w^2$, recovering the form used in Ref.~\onlinecite{Daas2020}.

%%%%%%%%%%%%%%%%%%%%%%
\section{Hydrogen molecule cation}
\label{sec:HMolCat}
%%%%%%%%%%%%%%%%%%%%%%

We now turn to the one-electron homonuclear diatomic \ce{H2+}, restricting our discussion to the ground state and its behavior in the dissociation limit.
In particular, we are interested in the possibility of spatially symmetry-broken UHF solutions for bond lengths greater than a (potentially) weight-dependent critical value,
forming an analogy to the Coulson--Fischer point in \ce{H2}. \cite{Coulson1949}
We employ the minimal basis set STO-3G containing two contracted $s$-type gaussian functions $\chi_L(\br)$ and $\chi_R(\br)$ centered on the left and 
right \ce{H} atoms respectively, with bond length $R$.
Again, the use of a minimal basis allows analytic expressions to be derived,
but the physics is similar in larger basis sets (see {\SupInt} for additional results).

The occupied and vacant (symmetry-pure) RHF orbitals form a delocalized orthogonal basis set
\begin{subequations}
\begin{align}
	\psi_1(\br) & = \frac{\chi_\tL(\br) + \chi_\tR(\br)}{\sqrt{2(1+S_{\tL \tR})}},
	\\
	\psi_2(\br) & = \frac{\chi_\tL(\br) - \chi_\tR(\br)}{\sqrt{2(1-S_{\tL \tR})}}, 
\end{align} 
\end{subequations} 
where $S_{\tL \tR} = \braket{\chi_\tL}{\chi_\tR}$ defines the overlap of the atomic orbitals at a given bond length.
The ground-state UHF orbitals can then be parametrized in terms of these orthogonal orbitals using a single rotation angle $\phi$ as\cite{SzaboBook}
\begin{subequations}
\begin{align}
	\psia_1(\br) & = \cos \phi \, \psi_1(\br) + \sin \phi \, \psi_2(\br),
	\\
	\psib_1(\br) & = \cos \phi \, \psi_1(\br) - \sin \phi \, \psi_2(\br),
\end{align}
\end{subequations}
while the corresponding virtual orbitals are
\begin{subequations}
\begin{align}
	\psia_2(\br) & = - \sin \phi \, \psi_1(\br) + \cos \phi \, \psi_2(\br),
	\\
	\psib_2(\br) & = + \sin \phi \, \psi_1(\br) + \cos \phi \, \psi_2(\br).
\end{align}
\end{subequations}
The symmetric RHF orbitals are recovered for $\phi = 0$, while $\phi = \pm \pi/4$ recovers spin-up and spin-down orbitals localized on opposite \ce{H} atoms.
The UHF ensemble energy is then given as
\begin{equation}
\begin{split}
	E_\UHF^w(\phi) 
	& = \cos^2 \phi \, h_{11} + \sin^2 \phi \, h_{22} 
	+ (1-w)w \Big[ \cos^4 \phi \, J_{11} 
	\\
	& + \sin^4 \phi \, J_{22} + 2 \cos^2 \phi \sin^2 \phi \, (J_{12} - 2 K_{12}) \Big],
\end{split}
\label{eq:EUHF_theta}
\end{equation}
where the one-electron, Coulomb, and exchange matrix elements in the orthogonal RHF basis are denoted
\begin{align}
	h_{ij} & = (\psi_i|h|\psi_j),	
	&
	J_{ij} & = (\psi_i \psi_i | \psi_j \psi_j),	
	&
	K_{ij} & = (\psi_i \psi_j | \psi_j \psi_i).
\end{align}

\begin{figure}
	\includegraphics[width=\columnwidth]{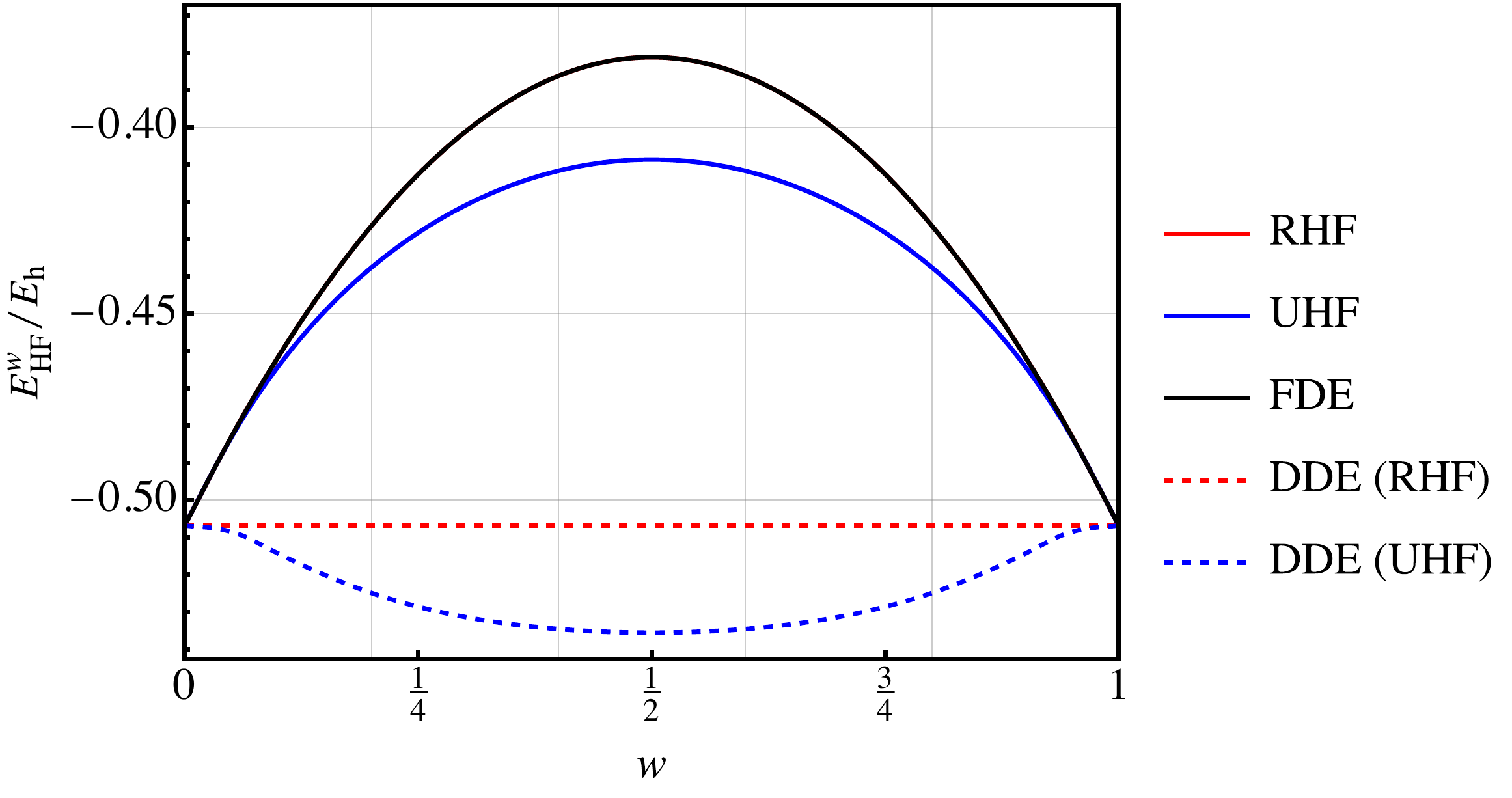}
\caption{RHF and UHF energies, functional-driven error (FDE), and density-driven error (DDE) as functions of $w$ for \ce{H2+} with fractional spins at $R= \SI{4}{\bohr}$ in the minimal STO-3G basis. 
In this case, the DDE for RHF is zero, while UHF stabilizes its energy by increasing the negative DDE.}
\label{fig:DDEH2}
\end{figure}

Differentiating Eq.~\eqref{eq:EUHF_theta} with respect to $\phi$ yields the ground-state RHF and UHF solutions as
\begin{subequations}
\begin{gather}
	\phi_\RHF = 0
	\\
	\cos^2 \phi_\UHF = \frac{h_{11} - h_{22} + 2w(1-w) (J_{12} - J_{22} - 2 K_{12}) }{-2w(1-w) (J_{11} - 2 J_{12} + J_{22} + 4 K_{22}) }.
\end{gather}
\end{subequations}
Analogously to \ce{H2}, the UHF solutions are spatially symmetry-broken, with the spin-up and spin-down electron densities localizing on opposite 
atoms, and do not necessarily exist for all ensemble weights or bond lengths.
The location of the (weight-dependent) Coulson--Fischer point can be identified by solving $\phi_\RHF = \phi_\UHF$, or equivalently
\begin{equation}
	\frac{h_{11} - h_{22} + 2w(1-w) (J_{12} - J_{22} - 2 K_{12}) }{-2w(1-w) (J_{11} - 2 J_{12} + J_{22} + 4 K_{22})} = 1.
\end{equation}
Therefore, in complete analogy with \ce{H2}, \cite{SzaboBook} there exists a (weight-dependent) critical bond-length $\Rc^w$ 
where, for $R > \Rc^ w$, it becomes energetically favorable for the spatial symmetry to be broken by localizing the 
high-spin and low-spin orbitals on opposite centers.
This symmetry breaking only occurs for $w \neq 0$ or $1$; at other ensemble weights, the UHF solution is
symmetric and equivalent to the RHF solution.
Like the \ce{H} atom, spatial separation of the high- and low-spin densities minimizes the {fractional-spin} error, leading to more accurate UHF 
energies, as illustrated in Fig.~\ref{fig:h2+}.
However, the energetic relaxation is more significant in \ce{H2+} than the \ce{H} atom as the two atomic centers increase the possible extent of spatial separation. 
Since the UHF energy is variational and the fractional spin error is positive, the energy and associated error are lowered by increasing the DDE (which is always negative).
For example, in Fig.~\ref{fig:DDEH2}, we show the minimal basis FDE and DDE at $R=\SI{4}{\bohr}$.
We find no DDE in RHF, while the DDE in UHF has maximum amplitude at $w=\frac{1}{2}$.

Figure~\ref{fig:h2+} also reveals that the critical bond-length $\Rc^w$ decreases as $w$ increases between 0 and $\frac{1}{2}$, reaching a minimum at $w = \frac{1}{2}$.
The shortest bond length for UHF symmetry breaking therefore occurs for $w = \frac{1}{2}$, which can be understood because the {fractional-spin} error is largest at this ensemble weight.
{The fractional-spin UHF binding curves with $w > 0$ fail to qualitatively describe the correct bound 
potential, instead predicting a shallow minimum and an unbound dissociation limit.
Furthermore, the UHF energy with $w = \frac{1}{2}$ has a similar
shape to the UHF energy of square \ce{H4^{2+}} shown in Fig.~2(d) of Ref.~\onlinecite{MoriSanchez2014}. 
This similarity can be understood by considering \ce{H4^{2+}} 
as two \ce{H2+} fragments each with $w = \frac{1}{2}$, suggesting that there is an intimate relationship
between the UHF fractional-spin error and the qualitative breakdown of the unrestricted Hartree--Fock energy.
We leave a detailed investigation of this relationship for future work.
}

%%%%%%%%%%%%%%%%%%%%%
%%%%%%%%%%%%%%%%%%%%%

Next, we consider the behavior of the RHF and UHF energies in the large-$R$ limit.
From Fig.~\ref{fig:h2+}, we see that the UHF energy decays towards the exact dissociation limit.
In this limit, the only remaining non-zero integrals are $h_{\tL\tL} = h_{\tR\tR} \equiv E_{\ce{H}}$ and 
$J_{\tL\tL} = J_{\tR\tR} = (\psi_{\ce{H}}\psi_{\ce{H}}|\psi_{\ce{H}}\psi_{\ce{H}})$, where $E_{\ce{H}}$ and $\psi_{\ce{H}}$ are the ground-state energy and wave function of an isolated \ce{H} atom;
all matrix elements involving different atomic centers decay to zero as $R$ grows.
The analytic RHF energy in the large-$R$ limit can then be derived as
\begin{equation}
	E_\RHF^w 
	\stackrel{R\to\infty}{=} 
    E_{\ce{H}} + \frac{w(1-w)}{2} (\psi_{\ce{H}}\psi_{\ce{H}}|\psi_{\ce{H}}\psi_{\ce{H}}).
\label{eq:RHFdissoc}
\end{equation}
Comparing Eqs.~\eqref{eq:RHFenergy} and \eqref{eq:RHFdissoc}, we find that the error in the RHF energy {of dissociated \ce{H_2^+}} is  
exactly half the fractional-spin error for the \ce{H} atom with high- and low-spin configurations of respective weights $(1-w)$ and $w$ (right panel of Fig.~\ref{fig:h2+}).
Note that this result is different to the two-electron $\ce{H2}$ molecule, where the dissociation error is \emph{equal} to the fractional-spin error in the \ce{H} 
atom.\cite{Cohen2008}
As a result, delocalization of the electron density over two dissociated atomic centers reduces the overall RHF {fractional-spin} error in one-electron models.

The reduction in the RHF {fractional-spin} error is even more pronounced for larger numbers of atomic centers as delocalizing the RHF electron density over
multiple sites further reduces the fictitious Coulomb repulsion between the spin-up and spin-down densities.
For example, it can be shown that the two-electron component of the RHF ensemble energy in a one-electron linear chain of $n$ equally-spaced \ce{H} atom 
decays as $\frac{w(1-w)}{n}$ at large $R$, while the UHF ensemble energy quickly decays to the correct energy.
By extension, these results suggest that static correlation errors in cationic molecules are likely to be less severe than their electron-dense anionic counterparts.

%%%%%%%%%%%%%%%%%%%%%%%%%%%%%%%%%%%%%%%%%%%%%%%%%%%%%%
\section{Concluding Remarks}
%%%%%%%%%%%%%%%%%%%%%%%%%%%%%%%%%%%%%%%%%%%%%%%%%%%%%%

% SUMMARISE
In this work, we have investigated the physical origins and consequences of 
the fractional-spin error in one-electron systems using the HF potential.
Building on the work of Cohen, Mori-Sanchez and Yang,\cite{Cohen2008,Cohen2008b,Cohen2012} we have explored
the variations in the fractional-spin error for the restricted, unrestricted, or the 
exact generalized HF formalisms.
Our results show that the fractional-spin error arises from an artificial Coulomb interaction between 
the high- and low-spin components of the one-electron density and, for all weights $w \neq 0$, $\frac{1}{2}$, and $1$, 
there exists an unrestricted solution in the \ce{H} atom that is lower in energy than the restricted one.
In the \ce{H} atom, this {fractional-spin} can also create additional solutions 
representing unphysical open-shell excited states, and a region of density 
depletion that mirrors the Coulomb hole in \ce{He}-like ions. 
These results further demonstrate that a fractional-spin electron behaves as a two-electron 
problem, and that electron correlation concepts fundamentally underpin
the physics of fractional-spin errors.
Finally, delocalization of the electron density reduces the 
fractional-spin error, and this effect is enhanced when the high- and 
low-spin electron densities can become spatially separated under an unrestricted 
approximation.

% PROPOSE
Overall, we have shown that the physics of one-electron systems is much
more subtle that it may first appear.
Our results demonstrate several new insights into the unusual properties of a fractional-spin electron 
within HF theory, and we also expect these properties
to occur in density-functional approximations as well. 
Looking forwards, we anticipate that these insights will inspire new directions for 
improving the performance of density functional approximations under the effects of 
strong static correlation.
In particular, the importance of capturing exchange-correlation effects between the 
high- and low-spin components of the electron density suggest that generalized 
extensions of DFT may provide a fruitful direction for future research.

%%%%%%%%%%%%%%%%%%%%%%%%%%%%%%%%%%%%%%%%%%%%%%%%%%%%%%%%%%%%%%
\section*{Supplementary Material}
%%%%%%%%%%%%%%%%%%%%%%%%%%%%%%%%%%%%%%%%%%%%%%%%%%%%%%%%%%%%%%
Included in the \SupInt{} are high-accuracy numerical data for 
the \ce{H} atom and \ce{H2+} cation that mirror the minimal basis results presented
in the main text.

%%%%%%%%%%%%%%%%%%%%%%%%%%%%%%%%%%%%%%%%%%%%%%%%%%%%%%%%%%%%%%
\begin{acknowledgements}
HGAB thanks New College, Oxford for support through the Astor Junior Research Fellowship.
CM thanks the Universit\'e Paul Sabatier (Toulouse, France) for a PhD scholarship.
TJD and PG-G were supported by the Netherlands Organisation for Scientific Research under Vici grant 724.017.001.
PFL thanks the European Research Council (ERC) under the European Union's Horizon 2020 research and innovation programme (grant agreement no.~863481) for financial support. 
\end{acknowledgements}

%%%%%%%%%%%%%%%%%%%%%%%%%%%%%%%%%%%%%%%%%%%%%
\appendix
%%%%%%%%%%%%%%%%%%%%%%%%%%%%%%%%%%%%%%%%%%%%%
\section{Generalized Hartree--Fock ensemble}
\label{apdx:GHF}
%%%%%%%%%%%%%%%%%%%%%%%%%%%%%%%%%%%%%%%%%%%%%
The aim of this Appendix is to demonstrate that a one-electron fractional-spin ensemble built from GHF densities must be independent 
of the ensemble weight, and thus always exact.
To achieve this, we attempt to build a two-state generalized ensemble density  in the same form of the UHF-based ensemble in  Eq.~\eqref{eq:rhoUHF}.
We find that any generalized ensemble of this form will give the exact energy for any ensemble weighting.

A GHF-based two-state ensemble can be seen as an extension of the UHF-based ensemble defined in Eq.~\eqref{eq:rhoUHF}.
The additional flexibility of the GHF formalism means that every orbital can include both a high-spin and low-spin component, giving
\begin{equation}
	\psi_{I}(\bx) = \psia_{I}(\br) \ket{\alpha} + \psib_{I}(\br) \ket{\beta},
\end{equation}
where $\bx =(\sigma,\br)$ is a composite coordinate gathering spin and spatial coordinates,
and the index $I=1,2$ labels the states belonging to the two-state ensemble.
In what follows, we will exploit the two-component spinor basis
\begin{align}
\ket{\alpha} & = \begin{pmatrix} 1 \\ 0 \end{pmatrix},
&
\ket{\beta} & = \begin{pmatrix} 0 \\ 1 \end{pmatrix},
\end{align}
such that 
\begin{equation}
\psi_{I} 
=
\begin{pmatrix}
\psia_{I}
\\
\psib_{I}
\end{pmatrix}.
\end{equation}
The corresponding GHF ensemble density is then
\begin{equation}
\rho_{\GHF}^w(\br) 
= (1-w) \, \rho_1(\br) + w \, \rho_2(\br),
\end{equation}
where the two-component density is defined as
\begin{equation}
\rho_{I}(\br) = 
\begin{pmatrix}
\rhoaa_{I}(\br) & \rhoab_{I}(\br) 
\\
\rhoba_{I}(\br)  & \rhobb_{I}(\br) 
\end{pmatrix},
\end{equation}
and $\rhossp_{I}(\br) = \psis_{I}(\br) \psisp_{I}(\br)$.
Similarly to the unrestricted ensemble, the only two-electron contributions to the ensemble energy arise from interactions
between the two densities. 
The corresponding GHF energy expression can then be derived as
\begin{equation}
\begin{split}
E_\GHF^w &= (1-w) h[\rho_{1}] + w h[\rho_{2}]
\\
&+ w(1-w) \sum_{\sigma\sigma'} \qty[ (\psi_1^{\sigma}\psi_1^{\sigma}| \psi_2^{\sigma'}\psi_2^{\sigma'}) - (\psi_1^{\sigma}\psi_2^{\sigma}| \psi_2^{\sigma'}\psi_1^{\sigma'}) ].
\end{split}
\label{eq:GHFenergy}
\end{equation}
From this expression, we see that the exact energy is recovered when the two individual densities, and their corresponding orbitals, are equivalent.
For any ensemble weight $w$, variational optimization of the ensemble density will therefore recover the exact energy $E_\GHF^w = E_0$ with 
$\psi_1 = \psi_2 = \psi_0$.
In other words, the two-state generalized ensemble introduces a redundant parametrization and, since the GHF representation of a one-electron system must always be exact, only a single GHF state is required.

Comparing Eq.~\eqref{eq:GHFenergy} with the UHF energy function Eq.~\eqref{eq:UHFenergy}, we see
that the fractional-spin error arises from the missing exchange interaction between high- and low-spin densities in the ensemble.
This missing term  in the restricted and unrestricted ensembles
leads to a {fractional-spin} error where the Coulomb interaction between the high- and low-spin components
is not sufficiently cancelled, leading to the static correlation error for fractional spins.

%\titou{Another useful geometrical representation of the GHF wave function is via the Bloch sphere \cite{Bloch1946} which is usually used to represent a two-level quantum mechanical system (or qubit). \cite{Nielsen2010}
%Indeed, the GHF wave function can be seen as a pure state of the Bloch sphere parametrized by the two orthogonal configurations $\ket{\alpha}$ and $\ket{\beta}$ corresponding to antipodal points on the Bloch sphere. 
%As a pure state, the GHF wave function lies on the Bloch sphere and is, by construction, an eigenfunction of the spin operator, while the RHF and UHF wave functions correspond to mixed states which lies inside the Bloch sphere.}

%%%%%%%%%%%%%%%%%%%%%%%%%%%%%%%%%%%%%%%%%%%%%
\section{Spin Expectation Values}
\label{apdx:S2equations}
%%%%%%%%%%%%%%%%%%%%%%%%%%%%%%%%%%%%%%%%%%%%%

To derive the spin-expectation values $\expval*{\cSx^2}$, $\expval*{\cSy^2}$, and $\expval*{\cSz^2}$ for the fractional-spin 
ensemble densities, we start with the second-quantized spin operators\cite{HelgakerBook}
\begin{subequations}
\begin{align}
\cSx &= \frac{1}{2} \sum_{p} \qty(a_{p \alpha}^\dagger a_{p \beta}^{\vphantom{\dagger}} + a_{p \beta}^\dagger a_{p \alpha}^{\vphantom{\dagger}} ),
\\
\cSy &= \frac{1}{2\mathrm{i}} \sum_{p} \qty(a_{p \beta}^\dagger a_{p \alpha}^{\vphantom{\dagger}} - a_{p \alpha}^\dagger a_{p \beta}^{\vphantom{\dagger}} ),
\\
\cSz &= \frac{1}{2} \sum_{p} \qty(a_{p \alpha}^\dagger a_{p \alpha}^{\vphantom{\dagger}} - a_{p \beta}^\dagger a_{p \beta}^{\vphantom{\dagger}} ).
\end{align}
\end{subequations}
Here, and in what follows, the indices $p$ and $q$ are used to denote an orthogonal spatial orbital basis used to construct the molecular orbitals.
Expanding the squared operators and 
considering the non-zero contractions\cite{ShavittBook}
allows the corresponding expectation values to be expressed using the two-component density matrices as
\begin{subequations}
\begin{align}
\begin{split}
\expval*{\cSx^2} & = 
\frac{1}{4}\sum_p \qty([n^{\alpha\alpha}]_{pp} + [n^{\beta\beta}]_{pp}) 
\\
& + \frac{1}{4}  \qty[\sum_p \qty([n^{\alpha\beta}]_{pp} + [n^{\beta\alpha}]_{pp}) ]^2 
\\
&-\frac{1}{2} \sum_{pq} \qty( [n^{\alpha\alpha}]_{pq} [n^{\beta\beta}]_{qp} + [n^{\beta\alpha}]_{pq} [n^{\alpha\beta}]_{qp} ),
\end{split}
\\
\begin{split}
\expval*{\cSy^2} & =
\frac{1}{4}\sum_p \qty([n^{\alpha\alpha}]_{pp} + [n^{\beta\beta}]_{pp}) 
\\
& - \frac{1}{4}  \qty[\sum_p \qty([n^{\alpha\beta}]_{pp} - [n^{\beta\alpha}]_{pp}) ]^2 
\\
&-\frac{1}{2} \sum_{pq} \qty( [n^{\alpha\alpha}]_{pq} [n^{\beta\beta}]_{qp} - [n^{\beta\alpha}]_{pq} [n^{\alpha\beta}]_{qp} ),
\end{split}
\\
\begin{split}
\expval*{\cSz^2} & = 
\frac{1}{4}\sum_p \qty([n^{\alpha\alpha}]_{pp} + [n^{\beta\beta}]_{pp} ) 
\\
& + \frac{1}{4}  \qty[\sum_p \qty([n^{\alpha\alpha}]_{pp} - [n^{\beta\beta}]_{pp}) ]^2 
\\
&-\frac{1}{4} \sum_{pq} \qty( [n^{\alpha\alpha}]_{pq} [n^{\alpha\alpha}]_{qp} + [n^{\beta\beta}]_{pq} [n^{\beta\beta}]_{qp} ) 
\\
&-\frac{1}{4} \sum_{pq} \qty( [n^{\alpha\beta}]_{pq} [n^{\beta\alpha}]_{qp} + [n^{\beta\alpha}]_{pq} [n^{\alpha\beta}]_{qp} ).
\end{split}
\end{align}
\end{subequations}
These formula are entirely generalized for any two-component electronic density matrix.

For the unrestricted ensemble density, where $n^{\alpha \beta} = n^{\beta \alpha} = 0$, these spin expectation values
 reduce to
\begin{subequations}
\begin{align}
\begin{split}
\expval*{\cSx^2} & = 
\frac{1}{4}\qty[(1-w) \Tr\,[n^{\alpha \alpha}]  + w \Tr\, [n^{\beta \beta}] ]
\\
& - \frac{(1-w)w}{2} \Tr\,[n^{\alpha \alpha} n^{\beta \beta}],
\end{split}
\\
\begin{split}
\expval*{\cSy^2} & = 
\frac{1}{4}\qty[(1-w) \Tr\,[n^{\alpha \alpha}]  + w \Tr\, [n^{\beta \beta}] ]
\\
& - \frac{(1-w)w}{2} \Tr\,[n^{\alpha \alpha} n^{\beta \beta}],
\end{split}
\\
\begin{split}
\expval*{\cSz^2} & = 
\frac{1}{4}\qty[(1-w) \Tr\,[n^{\alpha \alpha}]  + w \Tr\, [n^{\beta \beta}] ]
\\
& + \frac{1}{4}  \qty[(1-w) \Tr\,[n^{\alpha\alpha}] - w \Tr\,[n^{\beta\beta}] ]^2 
\\
& - \frac{1}{4} \qty[(1-w)^2 \Tr\,[n^{\alpha \alpha} n^{\alpha \alpha}] + w^2 \Tr\,[n^{\beta \beta} n^{\beta \beta}] ].
\end{split}
\end{align}
\end{subequations}
Further simplification can be achieved by noting that
individual state densities in the ensemble are normalized such that $\Tr\,[n^{\alpha \alpha}] = \Tr\, [n^{\beta \beta}] = 1$, and 
idempotent such that $n^{\alpha \alpha} n^{\alpha \alpha} = n^{\alpha \alpha}$ and $n^{\beta \beta}n^{\beta \beta}=n^{\beta \beta}$.
Furthermore, exploiting the invariance of the matrix trace  with respect to cyclic permutations allows 
 $\Tr\,[n^{\alpha \alpha} n^{\beta \beta}]$ to be expressed in terms of the spatial orbital between the occupied orbitals
as
\begin{equation}
\Tr\,[n^{\alpha \alpha} n^{\beta \beta}] = \Tr\,[S^{\alpha \beta} S^{\beta \alpha}],
\end{equation}
where the overlap elements are defined as $[S^{\alpha \beta}]_{ij} = \braket*{\psi_i^{\alpha}}{\psi_j^{\beta}}$.
Through these relationships, we recover the spin expectation values for the one-electron unrestricted fractional-spin ensemble as
\begin{subequations}
\begin{align}
\expval*{\cSx^2} &= 
\frac{1}{4}  - \frac{1}{2} w(1-w)\abs*{S^{\alpha \beta}}^2,
\\
\expval*{\cSy^2} &= 
\frac{1}{4}  - \frac{1}{2} w(1-w)\abs*{S^{\alpha \beta}}^2,
\\
\expval*{\cSz^2} &= 
\frac{1}{4}  - \frac{1}{2} w(1-w).
\end{align}
\end{subequations}

%%%%%%%%%%%%%%%%%%%%%%%%%%%%%%%%%%%%%%%%%%%%%
\section*{Data Availability}
%%%%%%%%%%%%%%%%%%%%%%%%%%%%%%%%%%%%%%%%%%%%%

The data that support the findings of this study are openly available in \textsc{zenodo}
at \href{https://doi.org/10.5281/zenodo.4765100}{https://doi.org/10.5281/zenodo.4765100}.

%%%%%%%%%%%%%%%%%%%%%%%%%%%%%%%%%%%%%%%%%%%%%%%%%%%%%%%%%%%%%%
\section*{References}
\bibliography{manuscript}

\begin{thebibliography}{96}%
\makeatletter
\providecommand \@ifxundefined [1]{%
 \@ifx{#1\undefined}
}%
\providecommand \@ifnum [1]{%
 \ifnum #1\expandafter \@firstoftwo
 \else \expandafter \@secondoftwo
 \fi
}%
\providecommand \@ifx [1]{%
 \ifx #1\expandafter \@firstoftwo
 \else \expandafter \@secondoftwo
 \fi
}%
\providecommand \natexlab [1]{#1}%
\providecommand \enquote  [1]{``#1''}%
\providecommand \bibnamefont  [1]{#1}%
\providecommand \bibfnamefont [1]{#1}%
\providecommand \citenamefont [1]{#1}%
\providecommand \href@noop [0]{\@secondoftwo}%
\providecommand \href [0]{\begingroup \@sanitize@url \@href}%
\providecommand \@href[1]{\@@startlink{#1}\@@href}%
\providecommand \@@href[1]{\endgroup#1\@@endlink}%
\providecommand \@sanitize@url [0]{\catcode `\\12\catcode `\$12\catcode
  `\&12\catcode `\#12\catcode `\^12\catcode `\_12\catcode `\%12\relax}%
\providecommand \@@startlink[1]{}%
\providecommand \@@endlink[0]{}%
\providecommand \url  [0]{\begingroup\@sanitize@url \@url }%
\providecommand \@url [1]{\endgroup\@href {#1}{\urlprefix }}%
\providecommand \urlprefix  [0]{URL }%
\providecommand \Eprint [0]{\href }%
\providecommand \doibase [0]{http://dx.doi.org/}%
\providecommand \selectlanguage [0]{\@gobble}%
\providecommand \bibinfo  [0]{\@secondoftwo}%
\providecommand \bibfield  [0]{\@secondoftwo}%
\providecommand \translation [1]{[#1]}%
\providecommand \BibitemOpen [0]{}%
\providecommand \bibitemStop [0]{}%
\providecommand \bibitemNoStop [0]{.\EOS\space}%
\providecommand \EOS [0]{\spacefactor3000\relax}%
\providecommand \BibitemShut  [1]{\csname bibitem#1\endcsname}%
\let\auto@bib@innerbib\@empty
%</preamble>
\bibitem [{\citenamefont {Fukutome}(1981)}]{Fukutome1981}%
  \BibitemOpen
  \bibfield  {author} {\bibinfo {author} {\bibfnamefont {H.}~\bibnamefont
  {Fukutome}},\ }\href {\doibase 10.1002/qua.560200502} {\bibfield  {journal}
  {\bibinfo  {journal} {Int. J. Quantum Chem.}\ }\textbf {\bibinfo {volume}
  {20}},\ \bibinfo {pages} {955} (\bibinfo {year} {1981})}\BibitemShut
  {NoStop}%
\bibitem [{\citenamefont {Sykja}\ and\ \citenamefont
  {Calais}(1982)}]{Sykja1982}%
  \BibitemOpen
  \bibfield  {author} {\bibinfo {author} {\bibfnamefont {B.}~\bibnamefont
  {Sykja}}\ and\ \bibinfo {author} {\bibfnamefont {J.~L.}\ \bibnamefont
  {Calais}},\ }\href {\doibase 10.1088/0022-3719/15/14/015} {\bibfield
  {journal} {\bibinfo  {journal} {J. Phys. C.: Solid State Phys.}\ }\textbf
  {\bibinfo {volume} {15}},\ \bibinfo {pages} {3079} (\bibinfo {year}
  {1982})}\BibitemShut {NoStop}%
\bibitem [{\citenamefont {Calais}(1985)}]{Calais1985}%
  \BibitemOpen
  \bibfield  {author} {\bibinfo {author} {\bibfnamefont {J.-L.}\ \bibnamefont
  {Calais}},\ }in\ \href {\doibase
  https://doi.org/10.1016/S0065-3276(08)60303-2} {\emph {\bibinfo {booktitle}
  {Adv. Quantum Chem.}}},\ Vol.~\bibinfo {volume} {17},\ \bibinfo {editor}
  {edited by\ \bibinfo {editor} {\bibfnamefont {P.-O.}\ \bibnamefont
  {L{\"o}wdin}}}\ (\bibinfo  {publisher} {Academic Press},\ \bibinfo {year}
  {1985})\ pp.\ \bibinfo {pages} {225--250}\BibitemShut {NoStop}%
\bibitem [{\citenamefont {L{\"o}wdin}\ and\ \citenamefont
  {Mayer}(1992)}]{Lowdin1992}%
  \BibitemOpen
  \bibfield  {author} {\bibinfo {author} {\bibfnamefont {P.-O.}\ \bibnamefont
  {L{\"o}wdin}}\ and\ \bibinfo {author} {\bibfnamefont {I.}~\bibnamefont
  {Mayer}}\ }(\bibinfo  {publisher} {Academic Press},\ \bibinfo {year} {1992})\
  pp.\ \bibinfo {pages} {79--114}\BibitemShut {NoStop}%
\bibitem [{\citenamefont {Mayer}\ and\ \citenamefont
  {L{\"{o}}wdin}(1993)}]{Mayer1993}%
  \BibitemOpen
  \bibfield  {author} {\bibinfo {author} {\bibfnamefont {I.}~\bibnamefont
  {Mayer}}\ and\ \bibinfo {author} {\bibfnamefont {P.-O.}\ \bibnamefont
  {L{\"{o}}wdin}},\ }\href {\doibase 10.1016/0009-2614(93)85341-K} {\bibfield
  {journal} {\bibinfo  {journal} {Chem.\ Phys.\ Lett.}\ }\textbf {\bibinfo
  {volume} {202}},\ \bibinfo {pages} {1} (\bibinfo {year} {1993})}\BibitemShut
  {NoStop}%
\bibitem [{\citenamefont {Hammes-Schiffer}\ and\ \citenamefont
  {Anderson}(1993)}]{HammesSchiffer1993}%
  \BibitemOpen
  \bibfield  {author} {\bibinfo {author} {\bibfnamefont {S.}~\bibnamefont
  {Hammes-Schiffer}}\ and\ \bibinfo {author} {\bibfnamefont {H.~C.}\
  \bibnamefont {Anderson}},\ }\href {\doibase 10.1063/1.465305} {\bibfield
  {journal} {\bibinfo  {journal} {J.\ Chem.\ Phys.}\ }\textbf {\bibinfo
  {volume} {99}},\ \bibinfo {pages} {1901} (\bibinfo {year}
  {1993})}\BibitemShut {NoStop}%
\bibitem [{\citenamefont {Stuber}\ and\ \citenamefont
  {Paldus}(2003)}]{Stuber2003}%
  \BibitemOpen
  \bibfield  {author} {\bibinfo {author} {\bibfnamefont {J.}~\bibnamefont
  {Stuber}}\ and\ \bibinfo {author} {\bibfnamefont {J.}~\bibnamefont
  {Paldus}},\ }\enquote {\bibinfo {title} {{Symmetry Breaking in the
  Independent Particle Model}},}\ in\ \href@noop {} {\emph {\bibinfo
  {booktitle} {Fundamental World of Quantum Chemistry: A Tribute to the Memory
  of Per-Olov L\"{o}wdin}}},\ Vol.~\bibinfo {volume} {1},\ \bibinfo {editor}
  {edited by\ \bibinfo {editor} {\bibfnamefont {E.~J.}\ \bibnamefont
  {Br\"{a}ndas}}\ and\ \bibinfo {editor} {\bibfnamefont {E.~S.}\ \bibnamefont
  {Kryachko}}}\ (\bibinfo  {publisher} {Kluwer Academic},\ \bibinfo {address}
  {Dordrecht},\ \bibinfo {year} {2003})\ p.~\bibinfo {pages} {67}\BibitemShut
  {NoStop}%
\bibitem [{\citenamefont {Jim{\'e}nez-Hoyos}, \citenamefont {Henderson},\ and\
  \citenamefont {Scuseria}(2011)}]{Jimenez-Hoyos2011}%
  \BibitemOpen
  \bibfield  {author} {\bibinfo {author} {\bibfnamefont {C.~A.}\ \bibnamefont
  {Jim{\'e}nez-Hoyos}}, \bibinfo {author} {\bibfnamefont {T.~M.}\ \bibnamefont
  {Henderson}}, \ and\ \bibinfo {author} {\bibfnamefont {G.~E.}\ \bibnamefont
  {Scuseria}},\ }\href {\doibase 10.1021/ct200345a} {\bibfield  {journal}
  {\bibinfo  {journal} {J. Chem. Theory Comput.}\ }\textbf {\bibinfo {volume}
  {7}},\ \bibinfo {pages} {2667} (\bibinfo {year} {2011})}\BibitemShut
  {NoStop}%
\bibitem [{\citenamefont {Small}, \citenamefont {Sundstrom},\ and\
  \citenamefont {Head-Gordon}(2015{\natexlab{a}})}]{Small2015a}%
  \BibitemOpen
  \bibfield  {author} {\bibinfo {author} {\bibfnamefont {D.~W.}\ \bibnamefont
  {Small}}, \bibinfo {author} {\bibfnamefont {E.~J.}\ \bibnamefont
  {Sundstrom}}, \ and\ \bibinfo {author} {\bibfnamefont {M.}~\bibnamefont
  {Head-Gordon}},\ }\href {\doibase 10.1063/1.4905120} {\bibfield  {journal}
  {\bibinfo  {journal} {J. Chem. Phys.}\ }\textbf {\bibinfo {volume} {142}},\
  \bibinfo {pages} {024104} (\bibinfo {year} {2015}{\natexlab{a}})}\BibitemShut
  {NoStop}%
\bibitem [{\citenamefont {Small}, \citenamefont {Sundstrom},\ and\
  \citenamefont {Head-Gordon}(2015{\natexlab{b}})}]{Small2015b}%
  \BibitemOpen
  \bibfield  {author} {\bibinfo {author} {\bibfnamefont {D.~W.}\ \bibnamefont
  {Small}}, \bibinfo {author} {\bibfnamefont {E.~J.}\ \bibnamefont
  {Sundstrom}}, \ and\ \bibinfo {author} {\bibfnamefont {M.}~\bibnamefont
  {Head-Gordon}},\ }\href {\doibase 10.1063/1.4913740} {\bibfield  {journal}
  {\bibinfo  {journal} {J. Chem. Phys.}\ }\textbf {\bibinfo {volume} {142}},\
  \bibinfo {pages} {094112} (\bibinfo {year} {2015}{\natexlab{b}})}\BibitemShut
  {NoStop}%
\bibitem [{\citenamefont {Goings}\ \emph {et~al.}(2015)\citenamefont {Goings},
  \citenamefont {Ding}, \citenamefont {Frisch},\ and\ \citenamefont
  {Li}}]{Goings2015}%
  \BibitemOpen
  \bibfield  {author} {\bibinfo {author} {\bibfnamefont {J.~J.}\ \bibnamefont
  {Goings}}, \bibinfo {author} {\bibfnamefont {F.}~\bibnamefont {Ding}},
  \bibinfo {author} {\bibfnamefont {M.~J.}\ \bibnamefont {Frisch}}, \ and\
  \bibinfo {author} {\bibfnamefont {X.}~\bibnamefont {Li}},\ }\href {\doibase
  10.1063/1.4918561} {\bibfield  {journal} {\bibinfo  {journal} {J. Chem.
  Phys.}\ }\textbf {\bibinfo {volume} {142}},\ \bibinfo {pages} {154109}
  (\bibinfo {year} {2015})}\BibitemShut {NoStop}%
\bibitem [{\citenamefont {Goings}, \citenamefont {Egidi},\ and\ \citenamefont
  {Li}(2018)}]{Goings2018}%
  \BibitemOpen
  \bibfield  {author} {\bibinfo {author} {\bibfnamefont {J.~J.}\ \bibnamefont
  {Goings}}, \bibinfo {author} {\bibfnamefont {F.}~\bibnamefont {Egidi}}, \
  and\ \bibinfo {author} {\bibfnamefont {X.}~\bibnamefont {Li}},\ }\href
  {\doibase 10.1002/qua.25398} {\bibfield  {journal} {\bibinfo  {journal} {Int.
  J. Quantum Chem.}\ }\textbf {\bibinfo {volume} {118}},\ \bibinfo {pages}
  {e25398} (\bibinfo {year} {2018})}\BibitemShut {NoStop}%
\bibitem [{\citenamefont {Henderson}, \citenamefont {Jim{\'e}nez-Hoyos},\ and\
  \citenamefont {Scuseria}(2018)}]{Henderson2018}%
  \BibitemOpen
  \bibfield  {author} {\bibinfo {author} {\bibfnamefont {T.~M.}\ \bibnamefont
  {Henderson}}, \bibinfo {author} {\bibfnamefont {C.~A.}\ \bibnamefont
  {Jim{\'e}nez-Hoyos}}, \ and\ \bibinfo {author} {\bibfnamefont {G.~E.}\
  \bibnamefont {Scuseria}},\ }\href {\doibase 10.1021/acs.jctc.7b01016}
  {\bibfield  {journal} {\bibinfo  {journal} {J. Chem. Theory Comput.}\
  }\textbf {\bibinfo {volume} {14}},\ \bibinfo {pages} {649} (\bibinfo {year}
  {2018})}\BibitemShut {NoStop}%
\bibitem [{\citenamefont {Jake}, \citenamefont {Henderson},\ and\ \citenamefont
  {Scuseria}(2018)}]{Jake2018}%
  \BibitemOpen
  \bibfield  {author} {\bibinfo {author} {\bibfnamefont {L.~C.}\ \bibnamefont
  {Jake}}, \bibinfo {author} {\bibfnamefont {T.~M.}\ \bibnamefont {Henderson}},
  \ and\ \bibinfo {author} {\bibfnamefont {G.~E.}\ \bibnamefont {Scuseria}},\
  }\href {\doibase 10.1063/1.5010929} {\bibfield  {journal} {\bibinfo
  {journal} {J. Chem. Phys.}\ }\textbf {\bibinfo {volume} {148}},\ \bibinfo
  {pages} {024109} (\bibinfo {year} {2018})}\BibitemShut {NoStop}%
\bibitem [{\citenamefont {Cohen}, \citenamefont {Mori-S\'{a}nchez},\ and\
  \citenamefont {Yang}(2008{\natexlab{a}})}]{Cohen2008}%
  \BibitemOpen
  \bibfield  {author} {\bibinfo {author} {\bibfnamefont {A.~J.}\ \bibnamefont
  {Cohen}}, \bibinfo {author} {\bibfnamefont {P.}~\bibnamefont
  {Mori-S\'{a}nchez}}, \ and\ \bibinfo {author} {\bibfnamefont
  {W.}~\bibnamefont {Yang}},\ }\href {\doibase 10.1063/1.2987202} {\bibfield
  {journal} {\bibinfo  {journal} {J.\ Chem.\ Phys.}\ }\textbf {\bibinfo
  {volume} {129}},\ \bibinfo {pages} {121104} (\bibinfo {year}
  {2008}{\natexlab{a}})}\BibitemShut {NoStop}%
\bibitem [{\citenamefont {Cohen}, \citenamefont {Mori-S\'{a}nchez},\ and\
  \citenamefont {Yang}(2008{\natexlab{b}})}]{Cohen2008b}%
  \BibitemOpen
  \bibfield  {author} {\bibinfo {author} {\bibfnamefont {A.~J.}\ \bibnamefont
  {Cohen}}, \bibinfo {author} {\bibfnamefont {P.}~\bibnamefont
  {Mori-S\'{a}nchez}}, \ and\ \bibinfo {author} {\bibfnamefont
  {W.}~\bibnamefont {Yang}},\ }\href {\doibase 10.1126/science.1158722}
  {\bibfield  {journal} {\bibinfo  {journal} {Science}\ }\textbf {\bibinfo
  {volume} {321}},\ \bibinfo {pages} {792} (\bibinfo {year}
  {2008}{\natexlab{b}})}\BibitemShut {NoStop}%
\bibitem [{\citenamefont {{Mori-S\'{a}nchez}}, \citenamefont {Cohen},\ and\
  \citenamefont {Yang}(2009)}]{MoriSanchez2009}%
  \BibitemOpen
  \bibfield  {author} {\bibinfo {author} {\bibfnamefont {P.}~\bibnamefont
  {{Mori-S\'{a}nchez}}}, \bibinfo {author} {\bibfnamefont {A.~J.}\ \bibnamefont
  {Cohen}}, \ and\ \bibinfo {author} {\bibfnamefont {W.}~\bibnamefont {Yang}},\
  }\href {\doibase 10.1103/PhysRevLett.102.066403} {\bibfield  {journal}
  {\bibinfo  {journal} {Phys.\ Rev.\ Lett.}\ }\textbf {\bibinfo {volume}
  {102}},\ \bibinfo {pages} {066403} (\bibinfo {year} {2009})}\BibitemShut
  {NoStop}%
\bibitem [{\citenamefont {Daas}\ \emph {et~al.}(2020)\citenamefont {Daas},
  \citenamefont {Grossi}, \citenamefont {Vuckovic}, \citenamefont {Musslimani},
  \citenamefont {Kooi}, \citenamefont {Seidl}, \citenamefont {Giesbertz},\ and\
  \citenamefont {Gori-Giorgi}}]{Daas2020}%
  \BibitemOpen
  \bibfield  {author} {\bibinfo {author} {\bibfnamefont {T.~J.}\ \bibnamefont
  {Daas}}, \bibinfo {author} {\bibfnamefont {J.}~\bibnamefont {Grossi}},
  \bibinfo {author} {\bibfnamefont {S.}~\bibnamefont {Vuckovic}}, \bibinfo
  {author} {\bibfnamefont {Z.~H.}\ \bibnamefont {Musslimani}}, \bibinfo
  {author} {\bibfnamefont {D.~P.}\ \bibnamefont {Kooi}}, \bibinfo {author}
  {\bibfnamefont {M.}~\bibnamefont {Seidl}}, \bibinfo {author} {\bibfnamefont
  {K.~J.~H.}\ \bibnamefont {Giesbertz}}, \ and\ \bibinfo {author}
  {\bibfnamefont {P.}~\bibnamefont {Gori-Giorgi}},\ }\href {\doibase
  10.1063/5.0029084} {\bibfield  {journal} {\bibinfo  {journal} {J. Chem.
  Phys.}\ }\textbf {\bibinfo {volume} {153}},\ \bibinfo {pages} {214112}
  (\bibinfo {year} {2020})}\BibitemShut {NoStop}%
\bibitem [{\citenamefont {Mussard}\ and\ \citenamefont
  {Toulouse}(2017)}]{Mussard2017}%
  \BibitemOpen
  \bibfield  {author} {\bibinfo {author} {\bibfnamefont {B.}~\bibnamefont
  {Mussard}}\ and\ \bibinfo {author} {\bibfnamefont {J.}~\bibnamefont
  {Toulouse}},\ }\href {\doibase 10.1080/00268976.2016.1213910} {\bibfield
  {journal} {\bibinfo  {journal} {Mol. Phys.}\ }\textbf {\bibinfo {volume}
  {115}},\ \bibinfo {pages} {161} (\bibinfo {year} {2017})}\BibitemShut
  {NoStop}%
\bibitem [{\citenamefont {He\ss{}elmann}\ and\ \citenamefont
  {G\"orling}(2011)}]{Hesselmann2011}%
  \BibitemOpen
  \bibfield  {author} {\bibinfo {author} {\bibfnamefont {A.}~\bibnamefont
  {He\ss{}elmann}}\ and\ \bibinfo {author} {\bibfnamefont {A.}~\bibnamefont
  {G\"orling}},\ }\href {\doibase 10.1103/PhysRevLett.106.093001} {\bibfield
  {journal} {\bibinfo  {journal} {Phys. Rev. Lett.}\ }\textbf {\bibinfo
  {volume} {106}},\ \bibinfo {pages} {093001} (\bibinfo {year}
  {2011})}\BibitemShut {NoStop}%
\bibitem [{\citenamefont {Erhard}, \citenamefont {Bleiziffer},\ and\
  \citenamefont {G\"orling}(2016)}]{Erhard2016}%
  \BibitemOpen
  \bibfield  {author} {\bibinfo {author} {\bibfnamefont {J.}~\bibnamefont
  {Erhard}}, \bibinfo {author} {\bibfnamefont {P.}~\bibnamefont {Bleiziffer}},
  \ and\ \bibinfo {author} {\bibfnamefont {A.}~\bibnamefont {G\"orling}},\
  }\href {\doibase 10.1103/PhysRevLett.117.143002} {\bibfield  {journal}
  {\bibinfo  {journal} {Phys. Rev. Lett.}\ }\textbf {\bibinfo {volume} {117}},\
  \bibinfo {pages} {143002} (\bibinfo {year} {2016})}\BibitemShut {NoStop}%
\bibitem [{\citenamefont {Chen}, \citenamefont {Friesecke},\ and\ \citenamefont
  {Mendl}(2014)}]{Chen2014}%
  \BibitemOpen
  \bibfield  {author} {\bibinfo {author} {\bibfnamefont {H.}~\bibnamefont
  {Chen}}, \bibinfo {author} {\bibfnamefont {G.}~\bibnamefont {Friesecke}}, \
  and\ \bibinfo {author} {\bibfnamefont {C.~B.}\ \bibnamefont {Mendl}},\ }\href
  {\doibase 10.1021/ct500586q} {\bibfield  {journal} {\bibinfo  {journal} {J.
  Chem. Theory Comput}\ }\textbf {\bibinfo {volume} {10}},\ \bibinfo {pages}
  {4360} (\bibinfo {year} {2014})}\BibitemShut {NoStop}%
\bibitem [{\citenamefont {Vuckovic}\ \emph {et~al.}(2015)\citenamefont
  {Vuckovic}, \citenamefont {Wagner}, \citenamefont {Mirtschink},\ and\
  \citenamefont {Gori-Giorgi}}]{Vuckovic2015}%
  \BibitemOpen
  \bibfield  {author} {\bibinfo {author} {\bibfnamefont {S.}~\bibnamefont
  {Vuckovic}}, \bibinfo {author} {\bibfnamefont {L.~O.}\ \bibnamefont
  {Wagner}}, \bibinfo {author} {\bibfnamefont {A.}~\bibnamefont {Mirtschink}},
  \ and\ \bibinfo {author} {\bibfnamefont {P.}~\bibnamefont {Gori-Giorgi}},\
  }\href {\doibase 10.1021/acs.jctc.5b00387} {\bibfield  {journal} {\bibinfo
  {journal} {J. Chem. Theory Comput.}\ }\textbf {\bibinfo {volume} {11}},\
  \bibinfo {pages} {3153} (\bibinfo {year} {2015})}\BibitemShut {NoStop}%
\bibitem [{\citenamefont {Vuckovic}\ and\ \citenamefont
  {Gori-Giorgi}(2017)}]{Vuckovic2017}%
  \BibitemOpen
  \bibfield  {author} {\bibinfo {author} {\bibfnamefont {S.}~\bibnamefont
  {Vuckovic}}\ and\ \bibinfo {author} {\bibfnamefont {P.}~\bibnamefont
  {Gori-Giorgi}},\ }\href {\doibase 10.1021/acs.jpclett.7b01113} {\bibfield
  {journal} {\bibinfo  {journal} {J. Phys. Chem. Lett.}\ }\textbf {\bibinfo
  {volume} {8}},\ \bibinfo {pages} {2799} (\bibinfo {year} {2017})}\BibitemShut
  {NoStop}%
\bibitem [{\citenamefont {Johnson}\ and\ \citenamefont
  {{Contrera-Garc\'{i}a}}(2011)}]{Johnson2011}%
  \BibitemOpen
  \bibfield  {author} {\bibinfo {author} {\bibfnamefont {E.~R.}\ \bibnamefont
  {Johnson}}\ and\ \bibinfo {author} {\bibfnamefont {J.}~\bibnamefont
  {{Contrera-Garc\'{i}a}}},\ }\href {\doibase 10.1063/1.3630117} {\bibfield
  {journal} {\bibinfo  {journal} {J.\ Chem.\ Phys.}\ }\textbf {\bibinfo
  {volume} {135}},\ \bibinfo {pages} {0881103} (\bibinfo {year}
  {2011})}\BibitemShut {NoStop}%
\bibitem [{\citenamefont {Cohen}, \citenamefont {Mori-S{\'a}nchez},\ and\
  \citenamefont {Yang}(2012)}]{Cohen2012}%
  \BibitemOpen
  \bibfield  {author} {\bibinfo {author} {\bibfnamefont {A.~J.}\ \bibnamefont
  {Cohen}}, \bibinfo {author} {\bibfnamefont {P.}~\bibnamefont
  {Mori-S{\'a}nchez}}, \ and\ \bibinfo {author} {\bibfnamefont
  {W.}~\bibnamefont {Yang}},\ }\href {\doibase 10.1021/cr200107z} {\bibfield
  {journal} {\bibinfo  {journal} {Chem. Rev.}\ }\textbf {\bibinfo {volume}
  {112}},\ \bibinfo {pages} {289} (\bibinfo {year} {2012})}\BibitemShut
  {NoStop}%
\bibitem [{\citenamefont {Su}, \citenamefont {Li},\ and\ \citenamefont
  {Yang}(2018)}]{Su2018}%
  \BibitemOpen
  \bibfield  {author} {\bibinfo {author} {\bibfnamefont {N.~Q.}\ \bibnamefont
  {Su}}, \bibinfo {author} {\bibfnamefont {C.}~\bibnamefont {Li}}, \ and\
  \bibinfo {author} {\bibfnamefont {W.}~\bibnamefont {Yang}},\ }\href {\doibase
  10.1073/pnas.1807095115} {\bibfield  {journal} {\bibinfo  {journal} {Proc.\
  Natl.\ Acad.\ Sci.\ U.S.A.}\ }\textbf {\bibinfo {volume} {115}},\ \bibinfo
  {pages} {9678} (\bibinfo {year} {2018})}\BibitemShut {NoStop}%
\bibitem [{\citenamefont {Phillips}, \citenamefont {Kananenka},\ and\
  \citenamefont {Zgid}(2015)}]{Phillips2015}%
  \BibitemOpen
  \bibfield  {author} {\bibinfo {author} {\bibfnamefont {J.~J.}\ \bibnamefont
  {Phillips}}, \bibinfo {author} {\bibfnamefont {A.~A.}\ \bibnamefont
  {Kananenka}}, \ and\ \bibinfo {author} {\bibfnamefont {D.}~\bibnamefont
  {Zgid}},\ }\href {\doibase 10.1063/1.4921259} {\bibfield  {journal} {\bibinfo
   {journal} {J.\ Chem.\ Phys.}\ }\textbf {\bibinfo {volume} {142}},\ \bibinfo
  {pages} {194108} (\bibinfo {year} {2015})}\BibitemShut {NoStop}%
\bibitem [{\citenamefont {Steinmann}\ and\ \citenamefont
  {Yang}(2013)}]{Steinmann2013}%
  \BibitemOpen
  \bibfield  {author} {\bibinfo {author} {\bibfnamefont {S.~N.}\ \bibnamefont
  {Steinmann}}\ and\ \bibinfo {author} {\bibfnamefont {W.}~\bibnamefont
  {Yang}},\ }\href {\doibase 10.1063/1.4817849} {\bibfield  {journal} {\bibinfo
   {journal} {J.\ Chem.\ Phys.}\ }\textbf {\bibinfo {volume} {139}},\ \bibinfo
  {pages} {074107} (\bibinfo {year} {2013})}\BibitemShut {NoStop}%
\bibitem [{\citenamefont {Cohen}, \citenamefont {{Mori-S\'{a}nchez}},\ and\
  \citenamefont {Yang}(2009)}]{Cohen2009}%
  \BibitemOpen
  \bibfield  {author} {\bibinfo {author} {\bibfnamefont {A.~J.}\ \bibnamefont
  {Cohen}}, \bibinfo {author} {\bibfnamefont {P.}~\bibnamefont
  {{Mori-S\'{a}nchez}}}, \ and\ \bibinfo {author} {\bibfnamefont
  {W.}~\bibnamefont {Yang}},\ }\href {\doibase 10.1021/ct8005419} {\bibfield
  {journal} {\bibinfo  {journal} {J.\ Chem.\ Theory Comput.}\ }\textbf
  {\bibinfo {volume} {5}},\ \bibinfo {pages} {786} (\bibinfo {year}
  {2009})}\BibitemShut {NoStop}%
\bibitem [{\citenamefont {Schipper}, \citenamefont {Gritsenko},\ and\
  \citenamefont {Baerends}(1998)}]{Schipper1998}%
  \BibitemOpen
  \bibfield  {author} {\bibinfo {author} {\bibfnamefont {P.}~\bibnamefont
  {Schipper}}, \bibinfo {author} {\bibfnamefont {O.}~\bibnamefont {Gritsenko}},
  \ and\ \bibinfo {author} {\bibfnamefont {E.}~\bibnamefont {Baerends}},\
  }\href {\doibase 10.1007/s002140050343} {\bibfield  {journal} {\bibinfo
  {journal} {Theor. Chem. Acc.}\ }\textbf {\bibinfo {volume} {99}},\ \bibinfo
  {pages} {329} (\bibinfo {year} {1998})}\BibitemShut {NoStop}%
\bibitem [{\citenamefont {Filatov}\ and\ \citenamefont
  {Shaik}(1999)}]{Filatov1999}%
  \BibitemOpen
  \bibfield  {author} {\bibinfo {author} {\bibfnamefont {M.}~\bibnamefont
  {Filatov}}\ and\ \bibinfo {author} {\bibfnamefont {S.}~\bibnamefont
  {Shaik}},\ }\href {\doibase https://doi.org/10.1016/S0009-2614(99)00336-X}
  {\bibfield  {journal} {\bibinfo  {journal} {Chem. Phys. Lett.}\ }\textbf
  {\bibinfo {volume} {304}},\ \bibinfo {pages} {429} (\bibinfo {year}
  {1999})}\BibitemShut {NoStop}%
\bibitem [{\citenamefont {Filatov}\ and\ \citenamefont
  {Shaik}(2000{\natexlab{a}})}]{Filatov2000a}%
  \BibitemOpen
  \bibfield  {author} {\bibinfo {author} {\bibfnamefont {M.}~\bibnamefont
  {Filatov}}\ and\ \bibinfo {author} {\bibfnamefont {S.}~\bibnamefont
  {Shaik}},\ }\href {\doibase 10.1021/jp0002289} {\bibfield  {journal}
  {\bibinfo  {journal} {J. Phys. Chem. A}\ }\textbf {\bibinfo {volume} {104}},\
  \bibinfo {pages} {6628} (\bibinfo {year} {2000}{\natexlab{a}})}\BibitemShut
  {NoStop}%
\bibitem [{\citenamefont {Filatov}\ and\ \citenamefont
  {Shaik}(2000{\natexlab{b}})}]{Filatov2000b}%
  \BibitemOpen
  \bibfield  {author} {\bibinfo {author} {\bibfnamefont {M.}~\bibnamefont
  {Filatov}}\ and\ \bibinfo {author} {\bibfnamefont {S.}~\bibnamefont
  {Shaik}},\ }\href {\doibase https://doi.org/10.1016/S0009-2614(00)01257-4}
  {\bibfield  {journal} {\bibinfo  {journal} {Chem. Phys. Lett.}\ }\textbf
  {\bibinfo {volume} {332}},\ \bibinfo {pages} {409} (\bibinfo {year}
  {2000}{\natexlab{b}})}\BibitemShut {NoStop}%
\bibitem [{\citenamefont {Ess}\ \emph {et~al.}(2011)\citenamefont {Ess},
  \citenamefont {Johnson}, \citenamefont {Hu},\ and\ \citenamefont
  {Yang}}]{Ess2011}%
  \BibitemOpen
  \bibfield  {author} {\bibinfo {author} {\bibfnamefont {D.~H.}\ \bibnamefont
  {Ess}}, \bibinfo {author} {\bibfnamefont {E.~R.}\ \bibnamefont {Johnson}},
  \bibinfo {author} {\bibfnamefont {X.}~\bibnamefont {Hu}}, \ and\ \bibinfo
  {author} {\bibfnamefont {W.}~\bibnamefont {Yang}},\ }\href {\doibase
  10.1021/jp109280y} {\bibfield  {journal} {\bibinfo  {journal} {J. Phys. Chem.
  A}\ }\textbf {\bibinfo {volume} {115}},\ \bibinfo {pages} {76} (\bibinfo
  {year} {2011})}\BibitemShut {NoStop}%
\bibitem [{\citenamefont {Chai}(2012)}]{Chai2012}%
  \BibitemOpen
  \bibfield  {author} {\bibinfo {author} {\bibfnamefont {J.-D.}\ \bibnamefont
  {Chai}},\ }\href {\doibase 10.1063/1.3703894} {\bibfield  {journal} {\bibinfo
   {journal} {J. Chem. Phys.}\ }\textbf {\bibinfo {volume} {136}},\ \bibinfo
  {pages} {154104} (\bibinfo {year} {2012})}\BibitemShut {NoStop}%
\bibitem [{\citenamefont {Filatov}(2015)}]{Filatov2015}%
  \BibitemOpen
  \bibfield  {author} {\bibinfo {author} {\bibfnamefont {M.}~\bibnamefont
  {Filatov}},\ }\href {\doibase https://doi.org/10.1002/wcms.1209} {\bibfield
  {journal} {\bibinfo  {journal} {{WIREs} Comput. Mol. Sci.}\ }\textbf
  {\bibinfo {volume} {5}},\ \bibinfo {pages} {146} (\bibinfo {year}
  {2015})}\BibitemShut {NoStop}%
\bibitem [{\citenamefont {Baerends}(2017)}]{Baerends2017}%
  \BibitemOpen
  \bibfield  {author} {\bibinfo {author} {\bibfnamefont {E.~J.}\ \bibnamefont
  {Baerends}},\ }\href {\doibase 10.1039/C7CP02123B} {\bibfield  {journal}
  {\bibinfo  {journal} {Phys. Chem. Chem. Phys.}\ }\textbf {\bibinfo {volume}
  {19}},\ \bibinfo {pages} {15639} (\bibinfo {year} {2017})}\BibitemShut
  {NoStop}%
\bibitem [{\citenamefont {Baerends}(2020)}]{Baerends2020}%
  \BibitemOpen
  \bibfield  {author} {\bibinfo {author} {\bibfnamefont {E.~J.}\ \bibnamefont
  {Baerends}},\ }\href {\doibase 10.1080/00268976.2019.1612955} {\bibfield
  {journal} {\bibinfo  {journal} {Mol. Phys.}\ }\textbf {\bibinfo {volume}
  {118}},\ \bibinfo {pages} {e1612955} (\bibinfo {year} {2020})}\BibitemShut
  {NoStop}%
\bibitem [{\citenamefont {Gidopoulos}, \citenamefont {Papaconstantinou},\ and\
  \citenamefont {Gross}(2002)}]{Gidopoulos2002}%
  \BibitemOpen
  \bibfield  {author} {\bibinfo {author} {\bibfnamefont {N.~I.}\ \bibnamefont
  {Gidopoulos}}, \bibinfo {author} {\bibfnamefont {P.~G.}\ \bibnamefont
  {Papaconstantinou}}, \ and\ \bibinfo {author} {\bibfnamefont {E.~K.~U.}\
  \bibnamefont {Gross}},\ }\href {\doibase 10.1103/PhysRevLett.88.033003}
  {\bibfield  {journal} {\bibinfo  {journal} {Phys. Rev. Lett.}\ }\textbf
  {\bibinfo {volume} {88}},\ \bibinfo {pages} {033003} (\bibinfo {year}
  {2002})}\BibitemShut {NoStop}%
\bibitem [{\citenamefont {{Mori-S\'{a}nchez}}\ and\ \citenamefont
  {Cohen}(2014)}]{MoriSanchez2014}%
  \BibitemOpen
  \bibfield  {author} {\bibinfo {author} {\bibfnamefont {P.}~\bibnamefont
  {{Mori-S\'{a}nchez}}}\ and\ \bibinfo {author} {\bibfnamefont {A.~J.}\
  \bibnamefont {Cohen}},\ }\href {\doibase 10.1063/1.4898860} {\bibfield
  {journal} {\bibinfo  {journal} {J.\ Chem.\ Phys.}\ }\textbf {\bibinfo
  {volume} {141}},\ \bibinfo {pages} {164124} (\bibinfo {year}
  {2014})}\BibitemShut {NoStop}%
\bibitem [{\citenamefont {Gordon}\ and\ \citenamefont
  {Truhlar}(1987)}]{Gordon1987}%
  \BibitemOpen
  \bibfield  {author} {\bibinfo {author} {\bibfnamefont {M.~S.}\ \bibnamefont
  {Gordon}}\ and\ \bibinfo {author} {\bibfnamefont {D.~G.}\ \bibnamefont
  {Truhlar}},\ }\href {\doibase 10.1007/BF00538477} {\bibfield  {journal}
  {\bibinfo  {journal} {Theor.\ Chem.\ Acc.}\ }\textbf {\bibinfo {volume}
  {71}},\ \bibinfo {pages} {1} (\bibinfo {year} {1987})}\BibitemShut {NoStop}%
\bibitem [{\citenamefont {Nobes}\ \emph {et~al.}(1991)\citenamefont {Nobes},
  \citenamefont {Moncrieff}, \citenamefont {Wong}, \citenamefont {Radom},
  \citenamefont {Gill},\ and\ \citenamefont {Pople}}]{Nobes1991}%
  \BibitemOpen
  \bibfield  {author} {\bibinfo {author} {\bibfnamefont {R.~H.}\ \bibnamefont
  {Nobes}}, \bibinfo {author} {\bibfnamefont {D.}~\bibnamefont {Moncrieff}},
  \bibinfo {author} {\bibfnamefont {M.~W.}\ \bibnamefont {Wong}}, \bibinfo
  {author} {\bibfnamefont {L.}~\bibnamefont {Radom}}, \bibinfo {author}
  {\bibfnamefont {P.~M.~W.}\ \bibnamefont {Gill}}, \ and\ \bibinfo {author}
  {\bibfnamefont {J.~A.}\ \bibnamefont {Pople}},\ }\href {\doibase
  10.1016/0009-2614(91)80204-B} {\bibfield  {journal} {\bibinfo  {journal}
  {Chem.\ Phys.\ Lett.}\ }\textbf {\bibinfo {volume} {182}},\ \bibinfo {pages}
  {216} (\bibinfo {year} {1991})}\BibitemShut {NoStop}%
\bibitem [{\citenamefont {Pople}(1971)}]{Pople1971}%
  \BibitemOpen
  \bibfield  {author} {\bibinfo {author} {\bibfnamefont {J.~A.}\ \bibnamefont
  {Pople}},\ }\href {\doibase 10.1002/qua.560050823} {\bibfield  {journal}
  {\bibinfo  {journal} {Int. J. Quantum Chem.}\ }\textbf {\bibinfo {volume}
  {5}},\ \bibinfo {pages} {175} (\bibinfo {year} {1971})}\BibitemShut {NoStop}%
\bibitem [{\citenamefont {Ostlund}(1972)}]{Ostlund1972}%
  \BibitemOpen
  \bibfield  {author} {\bibinfo {author} {\bibfnamefont {N.~S.}\ \bibnamefont
  {Ostlund}},\ }\href {\doibase 10.1063/1.1678695} {\bibfield  {journal}
  {\bibinfo  {journal} {J. Chem. Phys.}\ }\textbf {\bibinfo {volume} {57}},\
  \bibinfo {pages} {2994} (\bibinfo {year} {1972})}\BibitemShut {NoStop}%
\bibitem [{\citenamefont {Hiscock}\ and\ \citenamefont
  {Thom}(2014)}]{Hiscock2014}%
  \BibitemOpen
  \bibfield  {author} {\bibinfo {author} {\bibfnamefont {H.~G.}\ \bibnamefont
  {Hiscock}}\ and\ \bibinfo {author} {\bibfnamefont {A.~J.~W.}\ \bibnamefont
  {Thom}},\ }\href {\doibase 10.1021/ctc5007696} {\bibfield  {journal}
  {\bibinfo  {journal} {J. Chem. Theory Comput.}\ }\textbf {\bibinfo {volume}
  {10}},\ \bibinfo {pages} {4795} (\bibinfo {year} {2014})}\BibitemShut
  {NoStop}%
\bibitem [{\citenamefont {Burton}\ and\ \citenamefont
  {Thom}(2016)}]{Burton2016}%
  \BibitemOpen
  \bibfield  {author} {\bibinfo {author} {\bibfnamefont {H.~G.~A.}\
  \bibnamefont {Burton}}\ and\ \bibinfo {author} {\bibfnamefont {A.~J.~W.}\
  \bibnamefont {Thom}},\ }\href {\doibase acs.jctc.5b01005} {\bibfield
  {journal} {\bibinfo  {journal} {J. Chem. Theory Comput.}\ }\textbf {\bibinfo
  {volume} {12}},\ \bibinfo {pages} {167} (\bibinfo {year} {2016})}\BibitemShut
  {NoStop}%
\bibitem [{\citenamefont {Burton}, \citenamefont {Gross},\ and\ \citenamefont
  {Thom}(2018)}]{Burton2018}%
  \BibitemOpen
  \bibfield  {author} {\bibinfo {author} {\bibfnamefont {H.~G.~A.}\
  \bibnamefont {Burton}}, \bibinfo {author} {\bibfnamefont {M.}~\bibnamefont
  {Gross}}, \ and\ \bibinfo {author} {\bibfnamefont {A.~J.~W.}\ \bibnamefont
  {Thom}},\ }\href {\doibase 10.1021/acs.jctc.7b00980} {\bibfield  {journal}
  {\bibinfo  {journal} {J. Chem. Theory Comput.}\ }\textbf {\bibinfo {volume}
  {14}},\ \bibinfo {pages} {607} (\bibinfo {year} {2018})}\BibitemShut
  {NoStop}%
\bibitem [{\citenamefont {Coulson}\ and\ \citenamefont
  {Neilson}(1961)}]{Coulson1961}%
  \BibitemOpen
  \bibfield  {author} {\bibinfo {author} {\bibfnamefont {C.~A.}\ \bibnamefont
  {Coulson}}\ and\ \bibinfo {author} {\bibfnamefont {A.~H.}\ \bibnamefont
  {Neilson}},\ }\href {\doibase 10.1088/0370-1328/78/5/328} {\bibfield
  {journal} {\bibinfo  {journal} {Proc. Phys. Soc.}\ }\textbf {\bibinfo
  {volume} {78}},\ \bibinfo {pages} {831} (\bibinfo {year} {1961})}\BibitemShut
  {NoStop}%
\bibitem [{\citenamefont {Pearson}\ \emph {et~al.}(2009)\citenamefont
  {Pearson}, \citenamefont {Gill}, \citenamefont {Ugalde},\ and\ \citenamefont
  {Boyd}}]{Pearson2009}%
  \BibitemOpen
  \bibfield  {author} {\bibinfo {author} {\bibfnamefont {J.~K.}\ \bibnamefont
  {Pearson}}, \bibinfo {author} {\bibfnamefont {P.~M.}\ \bibnamefont {Gill}},
  \bibinfo {author} {\bibfnamefont {J.~M.}\ \bibnamefont {Ugalde}}, \ and\
  \bibinfo {author} {\bibfnamefont {R.~J.}\ \bibnamefont {Boyd}},\ }\href
  {\doibase 10.1080/00268970902740563} {\bibfield  {journal} {\bibinfo
  {journal} {Mol. Phys.}\ }\textbf {\bibinfo {volume} {107}},\ \bibinfo {pages}
  {1089} (\bibinfo {year} {2009})}\BibitemShut {NoStop}%
\bibitem [{\citenamefont {{Wolfram Research{,} Inc.}}()}]{Mathematica}%
  \BibitemOpen
  \bibfield  {author} {\bibinfo {author} {\bibnamefont {{Wolfram Research{,}
  Inc.}}},\ }\href@noop {} {\enquote {\bibinfo {title} {Mathematica, {V}ersion
  12.0},}\ }\bibinfo {note} {Champaign, IL, 2019}\BibitemShut {NoStop}%
\bibitem [{\citenamefont {Pastorczak}\ and\ \citenamefont
  {Pernal}(2014)}]{Pastorczak2014}%
  \BibitemOpen
  \bibfield  {author} {\bibinfo {author} {\bibfnamefont {E.}~\bibnamefont
  {Pastorczak}}\ and\ \bibinfo {author} {\bibfnamefont {K.}~\bibnamefont
  {Pernal}},\ }\href {\doibase 10.1063/1.4866998} {\bibfield  {journal}
  {\bibinfo  {journal} {J. Chem. Phys.}\ }\textbf {\bibinfo {volume} {140}},\
  \bibinfo {pages} {18A514} (\bibinfo {year} {2014})}\BibitemShut {NoStop}%
\bibitem [{\citenamefont {Alam}, \citenamefont {Knecht},\ and\ \citenamefont
  {Fromager}(2016)}]{Alam2016}%
  \BibitemOpen
  \bibfield  {author} {\bibinfo {author} {\bibfnamefont {M.~M.}\ \bibnamefont
  {Alam}}, \bibinfo {author} {\bibfnamefont {S.}~\bibnamefont {Knecht}}, \ and\
  \bibinfo {author} {\bibfnamefont {E.}~\bibnamefont {Fromager}},\ }\href
  {\doibase 10.1103/PhysRevA.94.012511} {\bibfield  {journal} {\bibinfo
  {journal} {Phys. Rev. A}\ }\textbf {\bibinfo {volume} {94}},\ \bibinfo
  {pages} {012511} (\bibinfo {year} {2016})}\BibitemShut {NoStop}%
\bibitem [{\citenamefont {Alam}\ \emph {et~al.}(2017)\citenamefont {Alam},
  \citenamefont {Deur}, \citenamefont {Knecht},\ and\ \citenamefont
  {Fromager}}]{Alam2017}%
  \BibitemOpen
  \bibfield  {author} {\bibinfo {author} {\bibfnamefont {M.~M.}\ \bibnamefont
  {Alam}}, \bibinfo {author} {\bibfnamefont {K.}~\bibnamefont {Deur}}, \bibinfo
  {author} {\bibfnamefont {S.}~\bibnamefont {Knecht}}, \ and\ \bibinfo {author}
  {\bibfnamefont {E.}~\bibnamefont {Fromager}},\ }\href {\doibase
  10.1063/1.4999825} {\bibfield  {journal} {\bibinfo  {journal} {J. Chem.
  Phys.}\ }\textbf {\bibinfo {volume} {147}},\ \bibinfo {pages} {204105}
  (\bibinfo {year} {2017})}\BibitemShut {NoStop}%
\bibitem [{\citenamefont {Gould}\ and\ \citenamefont
  {Pittalis}(2017)}]{Gould2017}%
  \BibitemOpen
  \bibfield  {author} {\bibinfo {author} {\bibfnamefont {T.}~\bibnamefont
  {Gould}}\ and\ \bibinfo {author} {\bibfnamefont {S.}~\bibnamefont
  {Pittalis}},\ }\href {\doibase 10.1103/PhysRevLett.119.243001} {\bibfield
  {journal} {\bibinfo  {journal} {Phys. Rev. Lett.}\ }\textbf {\bibinfo
  {volume} {119}},\ \bibinfo {pages} {243001} (\bibinfo {year}
  {2017})}\BibitemShut {NoStop}%
\bibitem [{\citenamefont {Loos}\ and\ \citenamefont
  {Fromager}(2020)}]{Loos2020}%
  \BibitemOpen
  \bibfield  {author} {\bibinfo {author} {\bibfnamefont {P.-F.}\ \bibnamefont
  {Loos}}\ and\ \bibinfo {author} {\bibfnamefont {E.}~\bibnamefont
  {Fromager}},\ }\href {\doibase 10.1063/5.0007388} {\bibfield  {journal}
  {\bibinfo  {journal} {J. Chem. Phys.}\ }\textbf {\bibinfo {volume} {152}},\
  \bibinfo {pages} {214101} (\bibinfo {year} {2020})}\BibitemShut {NoStop}%
\bibitem [{\citenamefont {Marut}\ \emph {et~al.}(2020)\citenamefont {Marut},
  \citenamefont {Senjean}, \citenamefont {Fromager},\ and\ \citenamefont
  {Loos}}]{Marut2020}%
  \BibitemOpen
  \bibfield  {author} {\bibinfo {author} {\bibfnamefont {C.}~\bibnamefont
  {Marut}}, \bibinfo {author} {\bibfnamefont {B.}~\bibnamefont {Senjean}},
  \bibinfo {author} {\bibfnamefont {E.}~\bibnamefont {Fromager}}, \ and\
  \bibinfo {author} {\bibfnamefont {P.-F.}\ \bibnamefont {Loos}},\ }\href
  {\doibase 10.1039/D0FD00059K} {\bibfield  {journal} {\bibinfo  {journal}
  {Faraday Discuss.}\ }\textbf {\bibinfo {volume} {224}},\ \bibinfo {pages}
  {402} (\bibinfo {year} {2020})}\BibitemShut {NoStop}%
\bibitem [{\citenamefont {Perdew}\ \emph {et~al.}(1982)\citenamefont {Perdew},
  \citenamefont {Parr}, \citenamefont {Levy},\ and\ \citenamefont
  {Balduz}}]{Perdew1982}%
  \BibitemOpen
  \bibfield  {author} {\bibinfo {author} {\bibfnamefont {J.~P.}\ \bibnamefont
  {Perdew}}, \bibinfo {author} {\bibfnamefont {R.~G.}\ \bibnamefont {Parr}},
  \bibinfo {author} {\bibfnamefont {M.}~\bibnamefont {Levy}}, \ and\ \bibinfo
  {author} {\bibfnamefont {J.~L.}\ \bibnamefont {Balduz}},\ }\href {\doibase
  10.1103/PhysRevLett.49.1691} {\bibfield  {journal} {\bibinfo  {journal}
  {Phys. Rev. Lett.}\ }\textbf {\bibinfo {volume} {49}},\ \bibinfo {pages}
  {1691} (\bibinfo {year} {1982})}\BibitemShut {NoStop}%
\bibitem [{\citenamefont {Gross}, \citenamefont {Oliveira},\ and\ \citenamefont
  {Kohn}(1988{\natexlab{a}})}]{Gross1988a}%
  \BibitemOpen
  \bibfield  {author} {\bibinfo {author} {\bibfnamefont {E.~K.~U.}\
  \bibnamefont {Gross}}, \bibinfo {author} {\bibfnamefont {L.~N.}\ \bibnamefont
  {Oliveira}}, \ and\ \bibinfo {author} {\bibfnamefont {W.}~\bibnamefont
  {Kohn}},\ }\href {\doibase 10.1103/PhysRevA.37.2805} {\bibfield  {journal}
  {\bibinfo  {journal} {Phys. Rev. A}\ }\textbf {\bibinfo {volume} {37}},\
  \bibinfo {pages} {2805} (\bibinfo {year} {1988}{\natexlab{a}})}\BibitemShut
  {NoStop}%
\bibitem [{\citenamefont {Gross}, \citenamefont {Oliveira},\ and\ \citenamefont
  {Kohn}(1988{\natexlab{b}})}]{Gross1988b}%
  \BibitemOpen
  \bibfield  {author} {\bibinfo {author} {\bibfnamefont {E.~K.~U.}\
  \bibnamefont {Gross}}, \bibinfo {author} {\bibfnamefont {L.~N.}\ \bibnamefont
  {Oliveira}}, \ and\ \bibinfo {author} {\bibfnamefont {W.}~\bibnamefont
  {Kohn}},\ }\href {\doibase 10.1103/PhysRevA.37.2809} {\bibfield  {journal}
  {\bibinfo  {journal} {Phys. Rev. A}\ }\textbf {\bibinfo {volume} {37}},\
  \bibinfo {pages} {2809} (\bibinfo {year} {1988}{\natexlab{b}})}\BibitemShut
  {NoStop}%
\bibitem [{\citenamefont {Oliveira}, \citenamefont {Gross},\ and\ \citenamefont
  {Kohn}(1988)}]{Oliveira1988}%
  \BibitemOpen
  \bibfield  {author} {\bibinfo {author} {\bibfnamefont {L.~N.}\ \bibnamefont
  {Oliveira}}, \bibinfo {author} {\bibfnamefont {E.~K.~U.}\ \bibnamefont
  {Gross}}, \ and\ \bibinfo {author} {\bibfnamefont {W.}~\bibnamefont {Kohn}},\
  }\href {\doibase 10.1103/PhysRevA.37.2821} {\bibfield  {journal} {\bibinfo
  {journal} {Phys. Rev. A}\ }\textbf {\bibinfo {volume} {37}},\ \bibinfo
  {pages} {2821} (\bibinfo {year} {1988})}\BibitemShut {NoStop}%
\bibitem [{\citenamefont {Senjean}\ and\ \citenamefont
  {Fromager}(2018)}]{Senjean2018}%
  \BibitemOpen
  \bibfield  {author} {\bibinfo {author} {\bibfnamefont {B.}~\bibnamefont
  {Senjean}}\ and\ \bibinfo {author} {\bibfnamefont {E.}~\bibnamefont
  {Fromager}},\ }\href {\doibase 10.1103/PhysRevA.98.022513} {\bibfield
  {journal} {\bibinfo  {journal} {Phys. Rev. A}\ }\textbf {\bibinfo {volume}
  {98}},\ \bibinfo {pages} {022513} (\bibinfo {year} {2018})}\BibitemShut
  {NoStop}%
\bibitem [{\citenamefont {Senjean}\ and\ \citenamefont
  {Fromager}(2020)}]{Senjean2019}%
  \BibitemOpen
  \bibfield  {author} {\bibinfo {author} {\bibfnamefont {B.}~\bibnamefont
  {Senjean}}\ and\ \bibinfo {author} {\bibfnamefont {E.}~\bibnamefont
  {Fromager}},\ }\href {\doibase https://doi.org/10.1002/qua.26190} {\bibfield
  {journal} {\bibinfo  {journal} {Int. J. Quantum Chem.}\ }\textbf {\bibinfo
  {volume} {120}},\ \bibinfo {pages} {e26190} (\bibinfo {year}
  {2020})}\BibitemShut {NoStop}%
\bibitem [{\citenamefont {Cohen}, \citenamefont {Mori-S\'anchez},\ and\
  \citenamefont {Yang}(2008)}]{Cohen2008c}%
  \BibitemOpen
  \bibfield  {author} {\bibinfo {author} {\bibfnamefont {A.~J.}\ \bibnamefont
  {Cohen}}, \bibinfo {author} {\bibfnamefont {P.}~\bibnamefont
  {Mori-S\'anchez}}, \ and\ \bibinfo {author} {\bibfnamefont {W.}~\bibnamefont
  {Yang}},\ }\href {\doibase 10.1103/PhysRevB.77.115123} {\bibfield  {journal}
  {\bibinfo  {journal} {Phys. Rev. B}\ }\textbf {\bibinfo {volume} {77}},\
  \bibinfo {pages} {115123} (\bibinfo {year} {2008})}\BibitemShut {NoStop}%
\bibitem [{\citenamefont {Kraisler}\ and\ \citenamefont
  {Kronik}(2013)}]{Kraisler2013}%
  \BibitemOpen
  \bibfield  {author} {\bibinfo {author} {\bibfnamefont {E.}~\bibnamefont
  {Kraisler}}\ and\ \bibinfo {author} {\bibfnamefont {L.}~\bibnamefont
  {Kronik}},\ }\href {\doibase 10.1103/PhysRevLett.110.126403} {\bibfield
  {journal} {\bibinfo  {journal} {Phys. Rev. Lett.}\ }\textbf {\bibinfo
  {volume} {110}},\ \bibinfo {pages} {126403} (\bibinfo {year}
  {2013})}\BibitemShut {NoStop}%
\bibitem [{\citenamefont {Kraisler}\ and\ \citenamefont
  {Kronik}(2014)}]{Kraisler2014}%
  \BibitemOpen
  \bibfield  {author} {\bibinfo {author} {\bibfnamefont {E.}~\bibnamefont
  {Kraisler}}\ and\ \bibinfo {author} {\bibfnamefont {L.}~\bibnamefont
  {Kronik}},\ }\href {\doibase 10.1063/1.4871462} {\bibfield  {journal}
  {\bibinfo  {journal} {J. Chem. Phys.}\ }\textbf {\bibinfo {volume} {140}},\
  \bibinfo {pages} {18A540} (\bibinfo {year} {2014})}\BibitemShut {NoStop}%
\bibitem [{\citenamefont {Cohen}, \citenamefont {Mori-S{\'a}nchez},\ and\
  \citenamefont {Yang}(2007)}]{Cohen2007}%
  \BibitemOpen
  \bibfield  {author} {\bibinfo {author} {\bibfnamefont {A.~J.}\ \bibnamefont
  {Cohen}}, \bibinfo {author} {\bibfnamefont {P.}~\bibnamefont
  {Mori-S{\'a}nchez}}, \ and\ \bibinfo {author} {\bibfnamefont
  {W.}~\bibnamefont {Yang}},\ }\href {\doibase 10.1063/1.2741248} {\bibfield
  {journal} {\bibinfo  {journal} {J. Chem. Phys.}\ }\textbf {\bibinfo {volume}
  {126}},\ \bibinfo {pages} {191109} (\bibinfo {year} {2007})}\BibitemShut
  {NoStop}%
\bibitem [{\citenamefont {Kraisler}\ and\ \citenamefont
  {Kronik}(2015)}]{Kraisler2015a}%
  \BibitemOpen
  \bibfield  {author} {\bibinfo {author} {\bibfnamefont {E.}~\bibnamefont
  {Kraisler}}\ and\ \bibinfo {author} {\bibfnamefont {L.}~\bibnamefont
  {Kronik}},\ }\href {\doibase 10.1103/PhysRevA.91.032504} {\bibfield
  {journal} {\bibinfo  {journal} {Phys. Rev. A}\ }\textbf {\bibinfo {volume}
  {91}},\ \bibinfo {pages} {032504} (\bibinfo {year} {2015})}\BibitemShut
  {NoStop}%
\bibitem [{\citenamefont {Kraisler}\ \emph {et~al.}(2015)\citenamefont
  {Kraisler}, \citenamefont {Schmidt}, \citenamefont {K{\"u}mmel},\ and\
  \citenamefont {Kronik}}]{Kraisler2015b}%
  \BibitemOpen
  \bibfield  {author} {\bibinfo {author} {\bibfnamefont {E.}~\bibnamefont
  {Kraisler}}, \bibinfo {author} {\bibfnamefont {T.}~\bibnamefont {Schmidt}},
  \bibinfo {author} {\bibfnamefont {S.}~\bibnamefont {K{\"u}mmel}}, \ and\
  \bibinfo {author} {\bibfnamefont {L.}~\bibnamefont {Kronik}},\ }\href
  {\doibase 10.1063/1.4930119} {\bibfield  {journal} {\bibinfo  {journal} {J.
  Chem. Phys.}\ }\textbf {\bibinfo {volume} {143}},\ \bibinfo {pages} {104105}
  (\bibinfo {year} {2015})}\BibitemShut {NoStop}%
\bibitem [{\citenamefont {Schmidt}\ and\ \citenamefont
  {Ruedenberg}(1979)}]{Schmidt1979}%
  \BibitemOpen
  \bibfield  {author} {\bibinfo {author} {\bibfnamefont {M.~W.}\ \bibnamefont
  {Schmidt}}\ and\ \bibinfo {author} {\bibfnamefont {K.}~\bibnamefont
  {Ruedenberg}},\ }\href {\doibase 10.1063/1.438165} {\bibfield  {journal}
  {\bibinfo  {journal} {J. Chem. Phys.}\ }\textbf {\bibinfo {volume} {71}},\
  \bibinfo {pages} {3951} (\bibinfo {year} {1979})}\BibitemShut {NoStop}%
\bibitem [{\citenamefont {Kim}, \citenamefont {Sim},\ and\ \citenamefont
  {Burke}(2013)}]{Kim2013}%
  \BibitemOpen
  \bibfield  {author} {\bibinfo {author} {\bibfnamefont {M.-C.}\ \bibnamefont
  {Kim}}, \bibinfo {author} {\bibfnamefont {E.}~\bibnamefont {Sim}}, \ and\
  \bibinfo {author} {\bibfnamefont {K.}~\bibnamefont {Burke}},\ }\href
  {\doibase 10.1103/PhysRevLett.111.073003} {\bibfield  {journal} {\bibinfo
  {journal} {Phys. Rev. Lett.}\ }\textbf {\bibinfo {volume} {111}},\ \bibinfo
  {pages} {073003} (\bibinfo {year} {2013})}\BibitemShut {NoStop}%
\bibitem [{\citenamefont {Vuckovic}\ \emph {et~al.}(2019)\citenamefont
  {Vuckovic}, \citenamefont {Song}, \citenamefont {Kozlowski}, \citenamefont
  {Sim},\ and\ \citenamefont {Burke}}]{Vuckovic2019}%
  \BibitemOpen
  \bibfield  {author} {\bibinfo {author} {\bibfnamefont {S.}~\bibnamefont
  {Vuckovic}}, \bibinfo {author} {\bibfnamefont {S.}~\bibnamefont {Song}},
  \bibinfo {author} {\bibfnamefont {J.}~\bibnamefont {Kozlowski}}, \bibinfo
  {author} {\bibfnamefont {E.}~\bibnamefont {Sim}}, \ and\ \bibinfo {author}
  {\bibfnamefont {K.}~\bibnamefont {Burke}},\ }\href {\doibase
  10.1021/acs.jctc.9b00826} {\bibfield  {journal} {\bibinfo  {journal} {J.
  Chem. Theory Comput.}\ }\textbf {\bibinfo {volume} {15}},\ \bibinfo {pages}
  {6636} (\bibinfo {year} {2019})}\BibitemShut {NoStop}%
\bibitem [{\citenamefont {Medvedev}\ \emph {et~al.}(2017)\citenamefont
  {Medvedev}, \citenamefont {Bushmarinov}, \citenamefont {Sun}, \citenamefont
  {Perdew},\ and\ \citenamefont {Lyssenko}}]{Medvedev2017}%
  \BibitemOpen
  \bibfield  {author} {\bibinfo {author} {\bibfnamefont {M.~G.}\ \bibnamefont
  {Medvedev}}, \bibinfo {author} {\bibfnamefont {I.~S.}\ \bibnamefont
  {Bushmarinov}}, \bibinfo {author} {\bibfnamefont {J.}~\bibnamefont {Sun}},
  \bibinfo {author} {\bibfnamefont {J.~P.}\ \bibnamefont {Perdew}}, \ and\
  \bibinfo {author} {\bibfnamefont {K.~A.}\ \bibnamefont {Lyssenko}},\ }\href
  {\doibase 10.1126/science.aah5975} {\bibfield  {journal} {\bibinfo  {journal}
  {Science}\ }\textbf {\bibinfo {volume} {355}},\ \bibinfo {pages} {49}
  (\bibinfo {year} {2017})}\BibitemShut {NoStop}%
\bibitem [{\citenamefont {Lykos}\ and\ \citenamefont
  {Pratt}(1963)}]{Lykos1963}%
  \BibitemOpen
  \bibfield  {author} {\bibinfo {author} {\bibfnamefont {P.}~\bibnamefont
  {Lykos}}\ and\ \bibinfo {author} {\bibfnamefont {G.~W.}\ \bibnamefont
  {Pratt}},\ }\href {\doibase 10.1103/RevModPhys.35.496} {\bibfield  {journal}
  {\bibinfo  {journal} {Rev. Mod. Phys.}\ }\textbf {\bibinfo {volume} {35}},\
  \bibinfo {pages} {496} (\bibinfo {year} {1963})}\BibitemShut {NoStop}%
\bibitem [{\citenamefont {Gori-Giorgi}\ and\ \citenamefont
  {Savin}(2005)}]{Gori-Giorgi2005}%
  \BibitemOpen
  \bibfield  {author} {\bibinfo {author} {\bibfnamefont {P.}~\bibnamefont
  {Gori-Giorgi}}\ and\ \bibinfo {author} {\bibfnamefont {A.}~\bibnamefont
  {Savin}},\ }\href@noop {} {\bibfield  {journal} {\bibinfo  {journal} {Phys.
  Rev. A}\ }\textbf {\bibinfo {volume} {{71}}},\ \bibinfo {pages} {032513}
  (\bibinfo {year} {2005})}\BibitemShut {NoStop}%
\bibitem [{\citenamefont {Burton}(2021)}]{Burton2021b}%
  \BibitemOpen
  \bibfield  {author} {\bibinfo {author} {\bibfnamefont {H.~G.~A.}\
  \bibnamefont {Burton}},\ }\href {\doibase 10.1063/5.0043105} {\bibfield
  {journal} {\bibinfo  {journal} {J.\ Chem.\ Phys.}\ }\textbf {\bibinfo
  {volume} {154}},\ \bibinfo {pages} {111103} (\bibinfo {year}
  {2021})}\BibitemShut {NoStop}%
\bibitem [{\citenamefont {Gill}\ \emph {et~al.}(2006)\citenamefont {Gill},
  \citenamefont {Crittenden}, \citenamefont {O'Neill},\ and\ \citenamefont
  {Besley}}]{Gill2006}%
  \BibitemOpen
  \bibfield  {author} {\bibinfo {author} {\bibfnamefont {P.~M.~W.}\
  \bibnamefont {Gill}}, \bibinfo {author} {\bibfnamefont {D.~L.}\ \bibnamefont
  {Crittenden}}, \bibinfo {author} {\bibfnamefont {D.~P.}\ \bibnamefont
  {O'Neill}}, \ and\ \bibinfo {author} {\bibfnamefont {N.~A.}\ \bibnamefont
  {Besley}},\ }\href {\doibase 10.1039/B511472A} {\bibfield  {journal}
  {\bibinfo  {journal} {Phys. Chem. Chem. Phys.}\ }\textbf {\bibinfo {volume}
  {8}},\ \bibinfo {pages} {15} (\bibinfo {year} {2006})}\BibitemShut {NoStop}%
\bibitem [{\citenamefont {Gori-Giorgi}, \citenamefont {Seidl},\ and\
  \citenamefont {Savin}(2008)}]{Gori-Giorgi2008}%
  \BibitemOpen
  \bibfield  {author} {\bibinfo {author} {\bibfnamefont {P.}~\bibnamefont
  {Gori-Giorgi}}, \bibinfo {author} {\bibfnamefont {M.}~\bibnamefont {Seidl}},
  \ and\ \bibinfo {author} {\bibfnamefont {A.}~\bibnamefont {Savin}},\ }\href
  {\doibase 10.1039/B803709B} {\bibfield  {journal} {\bibinfo  {journal} {Phys.
  Chem. Chem. Phys.}\ }\textbf {\bibinfo {volume} {10}},\ \bibinfo {pages}
  {3440} (\bibinfo {year} {2008})}\BibitemShut {NoStop}%
\bibitem [{\citenamefont {Per}, \citenamefont {Russo},\ and\ \citenamefont
  {Snook}(2009)}]{Per2009}%
  \BibitemOpen
  \bibfield  {author} {\bibinfo {author} {\bibfnamefont {M.~C.}\ \bibnamefont
  {Per}}, \bibinfo {author} {\bibfnamefont {S.~P.}\ \bibnamefont {Russo}}, \
  and\ \bibinfo {author} {\bibfnamefont {I.~K.}\ \bibnamefont {Snook}},\ }\href
  {\doibase 10.1063/1.3098353} {\bibfield  {journal} {\bibinfo  {journal} {J.
  Chem. Phys.}\ }\textbf {\bibinfo {volume} {130}},\ \bibinfo {pages} {134103}
  (\bibinfo {year} {2009})}\BibitemShut {NoStop}%
\bibitem [{\citenamefont {Via-Nadal}\ \emph {et~al.}(2019)\citenamefont
  {Via-Nadal}, \citenamefont {Rodr{\'\i}guez-Mayorga}, \citenamefont
  {Ramos-Cordoba},\ and\ \citenamefont {Matito}}]{Via-Nadal2019}%
  \BibitemOpen
  \bibfield  {author} {\bibinfo {author} {\bibfnamefont {M.}~\bibnamefont
  {Via-Nadal}}, \bibinfo {author} {\bibfnamefont {M.}~\bibnamefont
  {Rodr{\'\i}guez-Mayorga}}, \bibinfo {author} {\bibfnamefont {E.}~\bibnamefont
  {Ramos-Cordoba}}, \ and\ \bibinfo {author} {\bibfnamefont {E.}~\bibnamefont
  {Matito}},\ }\href {\doibase 10.1021/acs.jpclett.9b01376} {\bibfield
  {journal} {\bibinfo  {journal} {J. Phys. Chem. Lett.}\ }\textbf {\bibinfo
  {volume} {10}},\ \bibinfo {pages} {4032} (\bibinfo {year}
  {2019})}\BibitemShut {NoStop}%
\bibitem [{\citenamefont {Rodr{\'\i}guez-Mayorga}\ \emph
  {et~al.}(2019)\citenamefont {Rodr{\'\i}guez-Mayorga}, \citenamefont
  {Ramos-Cordoba}, \citenamefont {Lopez}, \citenamefont {Sol{\`a}},
  \citenamefont {Ugalde},\ and\ \citenamefont
  {Matito}}]{Rodriguez-Mayorga2019}%
  \BibitemOpen
  \bibfield  {author} {\bibinfo {author} {\bibfnamefont {M.}~\bibnamefont
  {Rodr{\'\i}guez-Mayorga}}, \bibinfo {author} {\bibfnamefont {E.}~\bibnamefont
  {Ramos-Cordoba}}, \bibinfo {author} {\bibfnamefont {X.}~\bibnamefont
  {Lopez}}, \bibinfo {author} {\bibfnamefont {M.}~\bibnamefont {Sol{\`a}}},
  \bibinfo {author} {\bibfnamefont {J.~M.}\ \bibnamefont {Ugalde}}, \ and\
  \bibinfo {author} {\bibfnamefont {E.}~\bibnamefont {Matito}},\ }\href
  {\doibase https://doi.org/10.1002/open.201800235} {\bibfield  {journal}
  {\bibinfo  {journal} {ChemistryOpen}\ }\textbf {\bibinfo {volume} {8}},\
  \bibinfo {pages} {411} (\bibinfo {year} {2019})}\BibitemShut {NoStop}%
\bibitem [{\citenamefont {Szabo}\ and\ \citenamefont
  {Ostlund}(1989)}]{SzaboBook}%
  \BibitemOpen
  \bibfield  {author} {\bibinfo {author} {\bibfnamefont {A.}~\bibnamefont
  {Szabo}}\ and\ \bibinfo {author} {\bibfnamefont {N.~S.}\ \bibnamefont
  {Ostlund}},\ }\href@noop {} {\emph {\bibinfo {title} {{Modern Quantum
  Chemistry}}}}\ (\bibinfo  {publisher} {Dover Publications Inc.},\ \bibinfo
  {year} {1989})\BibitemShut {NoStop}%
\bibitem [{\citenamefont {Coulson}\ and\ \citenamefont
  {Fischer}(1949)}]{Coulson1949}%
  \BibitemOpen
  \bibfield  {author} {\bibinfo {author} {\bibfnamefont {C.~A.}\ \bibnamefont
  {Coulson}}\ and\ \bibinfo {author} {\bibfnamefont {I.}~\bibnamefont
  {Fischer}},\ }\href {\doibase 10.1080/14786444908521726} {\bibfield
  {journal} {\bibinfo  {journal} {Philos.\ Mag.}\ }\textbf {\bibinfo {volume}
  {40}},\ \bibinfo {pages} {386} (\bibinfo {year} {1949})}\BibitemShut
  {NoStop}%
\bibitem [{\citenamefont {Burton}\ and\ \citenamefont
  {Wales}(2021)}]{Burton2021}%
  \BibitemOpen
  \bibfield  {author} {\bibinfo {author} {\bibfnamefont {H.~G.~A.}\
  \bibnamefont {Burton}}\ and\ \bibinfo {author} {\bibfnamefont {D.~J.}\
  \bibnamefont {Wales}},\ }\href {\doibase 10.1021/acs.jctc.0c00772} {\bibfield
   {journal} {\bibinfo  {journal} {J. Chem. Theory Comput.}\ }\textbf {\bibinfo
  {volume} {17}},\ \bibinfo {pages} {151} (\bibinfo {year} {2021})}\BibitemShut
  {NoStop}%
\bibitem [{\citenamefont {Andrews}\ \emph {et~al.}(1991)\citenamefont
  {Andrews}, \citenamefont {Jayatilaka}, \citenamefont {Bone}, \citenamefont
  {Handy},\ and\ \citenamefont {Amos}}]{Andrews1991}%
  \BibitemOpen
  \bibfield  {author} {\bibinfo {author} {\bibfnamefont {J.~S.}\ \bibnamefont
  {Andrews}}, \bibinfo {author} {\bibfnamefont {D.}~\bibnamefont {Jayatilaka}},
  \bibinfo {author} {\bibfnamefont {R.~G.~A.}\ \bibnamefont {Bone}}, \bibinfo
  {author} {\bibfnamefont {N.~C.}\ \bibnamefont {Handy}}, \ and\ \bibinfo
  {author} {\bibfnamefont {R.~D.}\ \bibnamefont {Amos}},\ }\href {\doibase
  10.1016/0009-2614(91)90405-X} {\bibfield  {journal} {\bibinfo  {journal}
  {Chem.\ Phys.\ Lett.}\ }\textbf {\bibinfo {volume} {183}},\ \bibinfo {pages}
  {423} (\bibinfo {year} {1991})}\BibitemShut {NoStop}%
\bibitem [{\citenamefont {Krylov}(2000)}]{Krylov2000}%
  \BibitemOpen
  \bibfield  {author} {\bibinfo {author} {\bibfnamefont {A.~I.}\ \bibnamefont
  {Krylov}},\ }\href {\doibase 10.1063/1.1308557} {\bibfield  {journal}
  {\bibinfo  {journal} {J. Chem. Phys.}\ }\textbf {\bibinfo {volume} {113}},\
  \bibinfo {pages} {6052} (\bibinfo {year} {2000})}\BibitemShut {NoStop}%
\bibitem [{\citenamefont {Lee}\ \emph {et~al.}(2018)\citenamefont {Lee},
  \citenamefont {Filatov}, \citenamefont {Lee},\ and\ \citenamefont
  {Choi}}]{Lee2018}%
  \BibitemOpen
  \bibfield  {author} {\bibinfo {author} {\bibfnamefont {S.}~\bibnamefont
  {Lee}}, \bibinfo {author} {\bibfnamefont {M.}~\bibnamefont {Filatov}},
  \bibinfo {author} {\bibfnamefont {S.}~\bibnamefont {Lee}}, \ and\ \bibinfo
  {author} {\bibfnamefont {C.~H.}\ \bibnamefont {Choi}},\ }\href {\doibase
  10.1063/1.5044202} {\bibfield  {journal} {\bibinfo  {journal} {J. Chem.
  Phys.}\ }\textbf {\bibinfo {volume} {149}},\ \bibinfo {pages} {104101}
  (\bibinfo {year} {2018})}\BibitemShut {NoStop}%
\bibitem [{\citenamefont {Casanova}\ and\ \citenamefont
  {Krylov}(2020)}]{Casanova2020}%
  \BibitemOpen
  \bibfield  {author} {\bibinfo {author} {\bibfnamefont {D.}~\bibnamefont
  {Casanova}}\ and\ \bibinfo {author} {\bibfnamefont {A.~I.}\ \bibnamefont
  {Krylov}},\ }\href {\doibase 10.1039/c9cp06507e} {\bibfield  {journal}
  {\bibinfo  {journal} {Phys. Chem. Chem. Phys.}\ }\textbf {\bibinfo {volume}
  {22}},\ \bibinfo {pages} {4326} (\bibinfo {year} {2020})}\BibitemShut
  {NoStop}%
\bibitem [{\citenamefont {Burton}, \citenamefont {Thom},\ and\ \citenamefont
  {Loos}(2019)}]{Burton2019a}%
  \BibitemOpen
  \bibfield  {author} {\bibinfo {author} {\bibfnamefont {H.~G.~A.}\
  \bibnamefont {Burton}}, \bibinfo {author} {\bibfnamefont {A.~J.~W.}\
  \bibnamefont {Thom}}, \ and\ \bibinfo {author} {\bibfnamefont {P.~F.}\
  \bibnamefont {Loos}},\ }\href {\doibase 10.1063/1.5085121} {\bibfield
  {journal} {\bibinfo  {journal} {J. Chem. Phys.}\ }\textbf {\bibinfo {volume}
  {150}},\ \bibinfo {pages} {041103} (\bibinfo {year} {2019})}\BibitemShut
  {NoStop}%
\bibitem [{\citenamefont {Marie}, \citenamefont {Burton},\ and\ \citenamefont
  {Loos}(2021)}]{Marie2021}%
  \BibitemOpen
  \bibfield  {author} {\bibinfo {author} {\bibfnamefont {A.}~\bibnamefont
  {Marie}}, \bibinfo {author} {\bibfnamefont {H.~G.~A.}\ \bibnamefont
  {Burton}}, \ and\ \bibinfo {author} {\bibfnamefont {P.~F.}\ \bibnamefont
  {Loos}},\ }\href {\doibase 10.1088/1361-648X/abe795} {\bibfield  {journal}
  {\bibinfo  {journal} {J. Phys.: Condens. Matter}\ ,\ \bibinfo {pages} {in
  press}} (\bibinfo {year} {2021})}\BibitemShut {NoStop}%
\bibitem [{\citenamefont {Stillinger}(2000)}]{Stillinger2000}%
  \BibitemOpen
  \bibfield  {author} {\bibinfo {author} {\bibfnamefont {F.~H.}\ \bibnamefont
  {Stillinger}},\ }\href {\doibase 10.1063/1.481608} {\bibfield  {journal}
  {\bibinfo  {journal} {J. Chem. Phys.}\ }\textbf {\bibinfo {volume} {112}},\
  \bibinfo {pages} {9711} (\bibinfo {year} {2000})}\BibitemShut {NoStop}%
\bibitem [{\citenamefont {Goodson}\ and\ \citenamefont
  {Sergeev}(2004)}]{Goodson2004}%
  \BibitemOpen
  \bibfield  {author} {\bibinfo {author} {\bibfnamefont {D.~Z.}\ \bibnamefont
  {Goodson}}\ and\ \bibinfo {author} {\bibfnamefont {A.~V.}\ \bibnamefont
  {Sergeev}},\ }\href {\doibase 10.1016/S0065-3276(04)47011-7} {\bibfield
  {journal} {\bibinfo  {journal} {Adv. Quant. Chem.}\ }\textbf {\bibinfo
  {volume} {47}},\ \bibinfo {pages} {193} (\bibinfo {year} {2004})}\BibitemShut
  {NoStop}%
\bibitem [{\citenamefont {Sergeev}\ \emph {et~al.}(2005)\citenamefont
  {Sergeev}, \citenamefont {Goodson}, \citenamefont {Wheeler},\ and\
  \citenamefont {Allen}}]{Sergeev2005}%
  \BibitemOpen
  \bibfield  {author} {\bibinfo {author} {\bibfnamefont {A.~V.}\ \bibnamefont
  {Sergeev}}, \bibinfo {author} {\bibfnamefont {D.~Z.}\ \bibnamefont
  {Goodson}}, \bibinfo {author} {\bibfnamefont {S.~E.}\ \bibnamefont
  {Wheeler}}, \ and\ \bibinfo {author} {\bibfnamefont {W.~D.}\ \bibnamefont
  {Allen}},\ }\href {\doibase 10.1063/1.1991854} {\bibfield  {journal}
  {\bibinfo  {journal} {J. Chem. Phys.}\ }\textbf {\bibinfo {volume} {123}},\
  \bibinfo {pages} {064105} (\bibinfo {year} {2005})}\BibitemShut {NoStop}%
\bibitem [{\citenamefont {Sergeev}\ and\ \citenamefont
  {Goodson}(2006)}]{Sergeev2006}%
  \BibitemOpen
  \bibfield  {author} {\bibinfo {author} {\bibfnamefont {A.~V.}\ \bibnamefont
  {Sergeev}}\ and\ \bibinfo {author} {\bibfnamefont {D.~Z.}\ \bibnamefont
  {Goodson}},\ }\href {\doibase 10.1063/1.2173989} {\bibfield  {journal}
  {\bibinfo  {journal} {J. Chem. Phys.}\ }\textbf {\bibinfo {volume} {124}},\
  \bibinfo {pages} {094111} (\bibinfo {year} {2006})}\BibitemShut {NoStop}%
\bibitem [{\citenamefont {Helgaker}, \citenamefont {J{\o}rgensen},\ and\
  \citenamefont {Olsen}(2000)}]{HelgakerBook}%
  \BibitemOpen
  \bibfield  {author} {\bibinfo {author} {\bibfnamefont {T.}~\bibnamefont
  {Helgaker}}, \bibinfo {author} {\bibfnamefont {P.}~\bibnamefont
  {J{\o}rgensen}}, \ and\ \bibinfo {author} {\bibfnamefont {J.}~\bibnamefont
  {Olsen}},\ }\href@noop {} {\emph {\bibinfo {title} {Molecular
  Electronic-Structure Theory}}}\ (\bibinfo  {publisher} {John Wiley {\&}
  Sons},\ \bibinfo {year} {2000})\BibitemShut {NoStop}%
\bibitem [{\citenamefont {Shavitt}\ and\ \citenamefont
  {Bartlett}(2009)}]{ShavittBook}%
  \BibitemOpen
  \bibfield  {author} {\bibinfo {author} {\bibfnamefont {I.}~\bibnamefont
  {Shavitt}}\ and\ \bibinfo {author} {\bibfnamefont {R.}~\bibnamefont
  {Bartlett}},\ }\href@noop {} {\emph {\bibinfo {title} {Many-Body Methods in
  Chemistry and Physics}}}\ (\bibinfo  {publisher} {Cambridge University
  Press},\ \bibinfo {year} {2009})\BibitemShut {NoStop}%
\end{thebibliography}%
%%%%%%%%%%%%%%%%%%%%%%%%%%%%%%%%%%%%%%%%%%%%%%%%%%%%%%%%%%%%%%

\end{document}